\documentclass[fleqn,usenatbib]{mnras}

\usepackage{amssymb}

\usepackage{newtxtext,newtxmath}

\usepackage[T1]{fontenc}
\usepackage{hyperref}

\DeclareRobustCommand{\VAN}[3]{#2}
\let\VANthebibliography\thebibliography
\def\thebibliography{\DeclareRobustCommand{\VAN}[3]{##3}\VANthebibliography}

\usepackage{graphicx}	
\usepackage{amsmath}	
\usepackage{amssymb}	

\usepackage[usenames,dvipsnames]{xcolor}
\newcommand{\Ion}[2]{#1{\,\scriptsize #2}}

\newcommand{\Porb}{\mbox{$P_{\rm orb}$}}
\newcommand{\Psh}{\mbox{$P_{\rm sh}$}}
\newcommand{\id}{\mbox{$\mathrm{d^{-1}}$}}

\title[Three poorly studied AM\,CVn stars]{Follow-up on three poorly studied AM\,CVn stars}

\author[Aungwerojwit et al.]{
Amornrat Aungwerojwit$^{1}$,
Boris T. G\"ansicke$^{2}$\thanks{E-mail: boris.gaensicke@gmail.com},
E. Breedt${^3}$,
S. Arjyotha$^{1}$\thanks{Deceased},
J.~J. Hermes$^{4}$, \newauthor
F.-J. Hambsch$^{5,6,7}$, 
A. Kumar$^{2,8}$,
S.~H. Ram\'irez$^{2}$,  
T.~G. Wilson$^{2}$,
V.~S. Dhillon$^{9,10}$, \newauthor
T.~R. Marsh$^{2}$\thanks{Deceased}, 
S. Poshyachinda$^{11}$, 
S. Scaringi$^{12,13}$,
J.~B. Haislip$^{14}$,
D.~E. Reichart$^{14}$,\\
$^{1}$Department of Physics, Faculty of Science, Naresuan University, Phitsanulok, 65000, Thailand\\
$^{2}$Department of Physics, University of Warwick, Coventry, CV4 7AL, UK\\
$^{3}$Institute of Astronomy, University of Cambridge, Cambridge, CB3 0HA, UK\\
$^{4}$Department of Astronomy, Boston University, 725 Commonwealth Ave., Boston, MA 02215, USA\\
$^{5}$Vereniging Voor Sterrenkunde (VVS), Oostmeers 122 C, 8000 Brugge, Belgium\\
$^{6}$Bundesdeutsche Arbeitsgemeinschaft f\"ur Ver\"anderliche Sterne, Munsterdamm 90, 12169 Berlin, Germany\\
$^{7}$AAVSO, 185 Alewife Brook Parkway, Suite 410, Cambridge, MA 02138, USA\\
$^{8}$Department of Physics, Royal Holloway, University of London, Egham Hill, Surrey, TW20 0EX, UK\\
$^{9}$Department of Physics and Astronomy, University of Sheffield, Sheffield S3 7RH, UK\\
$^{10}$Instituto de Astrof{\'i}sica de Canarias, E-38205 La Laguna, Tenerife, Spain\\
$^{11}$National Astronomical Research Institute of Thailand (Public Organization), Chiangmai, 50180, Thailand\\
$^{12}$Centre for Extragalactic Astronomy, Department of Physics, University of Durham, South Road, Durham DH1 3LE, UK\\
$^{13}$INAF-Osservatorio Astronomico di Capodimonte, Salita Moiariello 16, I-80131 Naples, Italy\\
$^{14}$Department of Physics and Astronomy, University of North Carolina at Chapel Hill, Chapel Hill, NC 27599, USA\\
}

\date{Accepted XXX. Received YYY; in original form ZZZ}

\pubyear{2021}

\begin{document}
\label{firstpage}
\pagerange{\pageref{firstpage}--\pageref{lastpage}}
\maketitle

\begin{abstract}
We report follow-up observations of three poorly studied AM\,CVn-type binaries: CRTS CSS150211 J091017$-$200813, NSV\,1440, and SDSS\,J183131.63+420220.2. Analysing time-series photometry obtained with a range of ground-based facilities as well as with \textit{TESS}, we determine the superhump period of CRTS\,J0910$-$2008 as $\Psh=29.700\pm0.004$\,min and the orbital period of NSV\,1440 as $\Porb=36.56\pm0.03$\,min. We also confirm a photometric period of $P=23.026\pm0.097$\,min in SDSS\,J1831+4202, which is most likely the superhump period. We also report the first optical spectroscopy of CRTS\,J0910$-$2008 and NSV\,1440 which unambiguously confirms both as AM\,CVn systems. We briefly discuss the distribution in the Hertzsprung-Russell diagram of the currently known sample of 63 AM\,CVn stars with known periods and \textit{Gaia} data. 
\end{abstract}


\section{Introduction} 
AM\,CVn-type stars are short-period ($\Porb\simeq5-68$\,min), helium-dominated binaries in which a white dwarf accretes material from a close companion through Roche lobe overflow. Their compact configuration implies that the companion itself must be an evolved star, likely another white dwarf or a semi-degenerate helium star. They are expected to be bright sources of low-frequency gravitational waves \citep{nelemansetal04-1, amaro-seoaneetal12-1} and hence some of the first sources to be detected by the \textit{LISA} Gravitational Wave Observatory \citep{koroletal17-1, kupferetal18-1}. AM\,CVn stars may also be the sources of helium novae \citep{katoetal08-1, woudtetal09-1}, Type\,Ia supernovae \citep{shen+bildsten14-1} and the sub-luminous `.Ia' explosions \citep{bildstenetal07-1}. For a detailed review and a summary of the known population's properties, see \citet{solheim10-1} and \citet{ramsayetal18-1}, respectively. 

In all but the most compact of these binaries \citep{marshetal04-1}, accretion occurs via an accretion disc. Just as is the case in the hydrogen-dominated cataclysmic variables (CVs) \citep{osaki74-1}, the helium discs of AM\,CVn stars are subject to a thermal instability which can cause quasi-regular outbursts of several magnitudes \citep{ramsayetal12-2, kotkoetal12-1}. As a result, AM\,CVn systems are detected as transients in large area photometric surveys including the Catalina Real-time Transient Survey (CRTS; \citealt{drakeetal09-1}), the All-Sky Automated Survey for Supernovae (ASAS-SN; \citealt{shappeeetal14-1, kochaneketal17-1}), Zwicky Transient Facility (ZTF; \citealt{bellmetal19-1, mascietal19-1}) and the Gravitational-wave Optical Transient Observer (GOTO; \citealt{steeghsetal22-1}). The dwarf nova outbursts of AM\,CVn are generally not distinguishable from those of the more common hydrogen CVs, and additional follow-up is usually required to confirm the AM\,CVn nature \citep{levitanetal11-1, vanroesteletal21-1, khaliletal24-1}. 
 
AM\,CVn stars with $\Porb\lesssim18$\,min have stable, hot accretion discs and do not display outbursts and systems with $\Porb\gtrsim45$\,min have such low mass transfer rate that they rarely outburst (e.g. \citealt{levitanetal15-1}, even though the outburst properties are somewhat more complex, see \citealt{duffyetal21-1}). Discovering these systems is more difficult and requires either the detection of eclipses \citep{vanroesteletal22-1}, orbital variability \citep{smak67-1}, spectra dominated by helium lines \citep{greenstein+matthews57-1, carteretal13-1} or X-ray emission \citep{rodriguezetal23-1}. The serendipitous discovery of bright  ($G\simeq14.3$) short-period ($\Porb\simeq23$\,min) AM\,CVn systems from \textit{TESS} photometry demonstrates that the current census of AM\,CVn systems is still strongly biased and incomplete \citep{greenetal24-1}.

\begin{table*}
\caption{Basic properties of the three AM\,CVn stars analysed in this paper. \label{t-properties}.}
\setlength{\tabcolsep}{0.95ex}
\begin{flushleft}
\begin{tabular}{lccccc}
\hline\noalign{\smallskip}
Object & RA (IRCS 2000) & Dec (IRCS 2000)  & \textit{Gaia} DR3 \texttt{source\_id} & Parallax (mas) &  Typical magnitude (mag) \\
\hline\noalign{\smallskip}
CRTS CSS150211 J091017$-$200813 & 09:10:17.45   & $-$20:08:12.5 & 5679545150277507328 & $1.35\pm0.14$ & $G=19.4$ \\
NSV\,1440                       & 03:55:17.97   & $-$82:26:11.3 & 4616023664815517440 & $3.08\pm0.08$ & $G=18.5$ \\
SDSS\,J183131.63+420220.2       & 18:31:31.63   &   +42:02:20.2 & 2111270721441518208 & $0.71\pm0.07$ & $G=17.5$ \\

\hline
\end{tabular}
\end{flushleft}
\end{table*}

\begin{figure*}
\centerline{\includegraphics[width=0.85\textwidth]{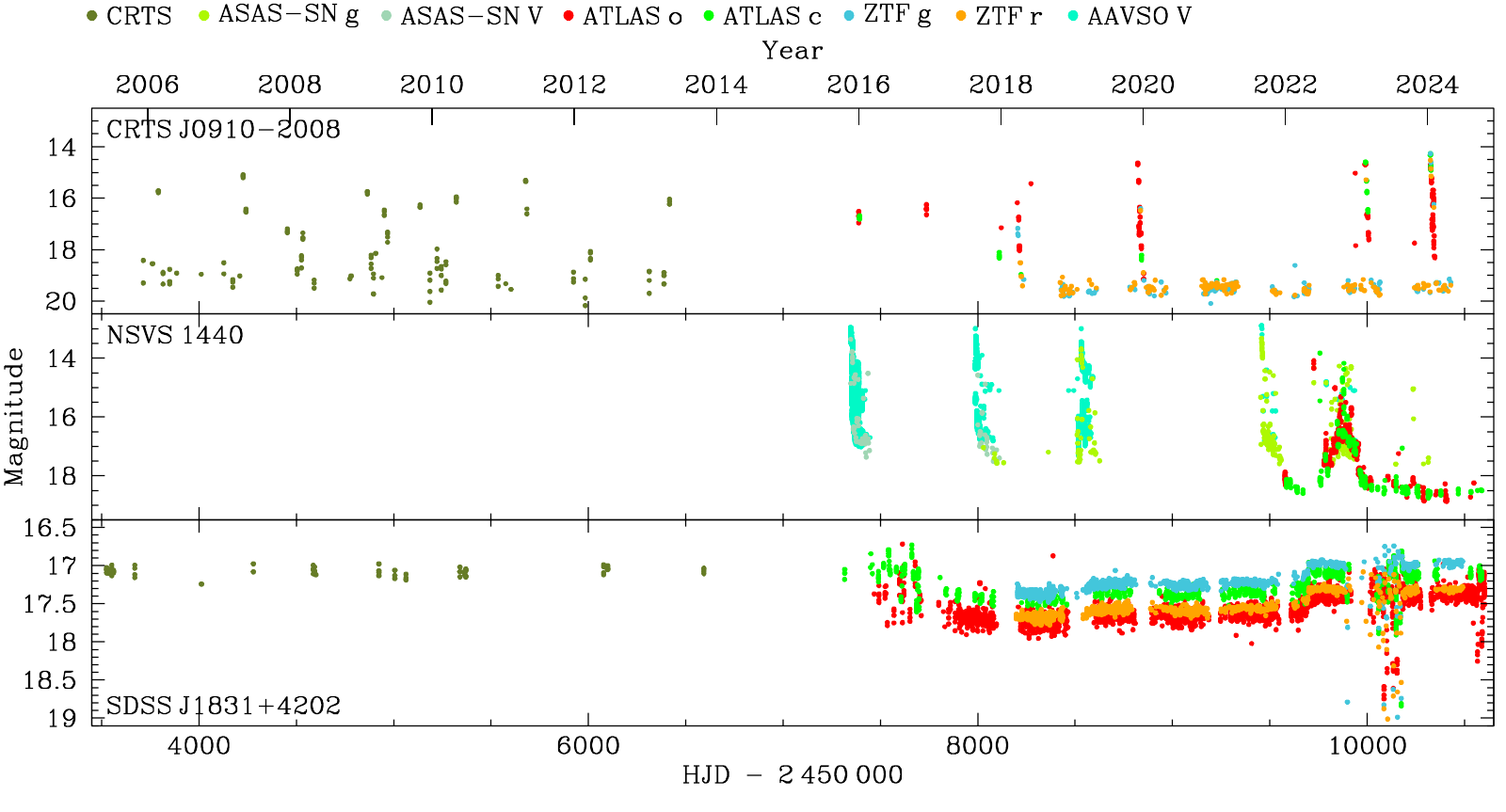}}
\caption{\label{f-surveylcs} Long-term archival light curves of the three AM\,CVn stars discussed in this paper. CRTS\,J0910$-$2008 and NSV\,1440 display frequent outbursts. SDSS\,J1831+4202 spends most of the time in a stable high state with some changes in its long-term brightness, but undergoes a short phase during which the accretion disc becomes thermally unstable ($\mathrm{HJD}\simeq2\,457\,500$ and $\mathrm{HJD}\simeq2\,460\,000$).}
\end{figure*}

As AM\,CVn systems are both less common and intrinsically less luminous (because of their smaller accretion discs) than hydrogen-rich CVs, the majority of the known systems are faint:  \citet{green23-1} maintains a catalogue of AM\,CVn stars\footnote{https://zenodo.org/records/12672892}, and the median apparent magnitude of the 107 confirmed systems is $G\simeq19.5$. Consequently, a detailed characterisation of this population is challenging, in particular orbital phase-resolved spectroscopy, even with large-aperture telescopes. Fortunately, the orbital period of AM\,CVn stars can often be estimated photometrically from a phenomenon known as superhumps, a dynamical interaction of the outer edge of the accretion disc with the companion star \citep{whitehurst+king91-1} resulting in brightness variations on a period with a few per cent longer than the orbital one \citep{pattersonetal05-3, katoetal09-1}. 

Here we report time-series photometry of three AM\,CVn candidates obtained using a variety of small to medium-size telescopes (0.4--2.4\,m). Sections~\ref{sec-obs} and \ref{sec-analysis} describe our follow-up observation and analysis of the data. We discuss the three new binaries in the context of the known population in Section~\ref{sec-discussion}. 

\section{The three AM\,CVn systems} 
  
\subsection{CRTS CSS150211 J091017$-$200813} 
We obtained the first observations of CRTS CSS150211 J091017$-$200813 (hereafter CRTS\,J0910$-$2008, see Table\,\ref{t-properties}) in response to an outburst detected by the Catalina Sky Survey (CSS) telescope of CRTS on 2015 February 11 \citep{drakeetal09-1}. The system, typically at $V\simeq19.5$\,mag in quiescence, was detected at $V=16.1$ during this outburst. This is the only outburst of this system observed by CSS, but upon querying the CRTS archive, we discovered that the CRTS Siding Springs Survey (SSS) telescope observed many previous outbursts of this system, the brightest of which was measured to be $V=15.08$. The full CRTS light curve covered more than 12 years is shown in Fig.\,\ref{f-surveylcs}. We detected another outburst of this object on 2018 March 18, while observing it using the 2.4-m Thai National Telescope (TNT). The object reached $KG5\simeq14.0$ during our observations\footnote{The $KG5$ filter is a broad filter which includes most of the SDSS $u'$, $g'$ and $r'$ band-passes. See \citet{hardyetal17-1} for the transmission curve and further details.}, but it was not observed during this time by the transient surveys.

\subsection{ASASSN-15sz / NSV\,1440}
The ASAS-SN survey alerted on a potential cataclysmic variable, ASASSN-15sz (see Table\,\ref{t-properties}), on 2015 November 20. The discovery outburst reached $V=12.84$ (Fig.\,\ref{f-surveylcs}), but the target is below the ASAS-SN detection limit ($V\lesssim17$) in quiescence. It was however detected by Gaia at $G=18.52$ \citep[DR1,][]{gaiaetal18-1}. The International Variable Star index \citep[VSX,][]{watsonetal16-1} revealed that the object had previously been identified as a variable star, although its nature was not determined at the time. The name originally assigned to the variable, NSV\,1440\footnote{At the time of publishing, Simbad reports the wrong object for NSV\,1440.}, has been widely adopted for this star, and we follow that convention here. The 2015 outburst, as well as another in 2017, was studied in detail by \citet{isogaietal19-1}. They showed that NSV\,1440 was the first AM\,CVn system to display the double superoutburst characteristic that is normally seen in hydrogen dwarf novae of WZ\,Sge-type\footnote{WZ\,Sge stars are evolved hydrogen cataclysmic variables with large amplitude, but infrequent, outbursts. See \citet{kato15-1} for a detailed review.}.


\subsection{SDSS\,J183131.63+420220.2}
SDSS\,J183131.63+420220.2 (hereafter SDSS\,J1831+4202, see Table\,\ref{t-properties}) was identified by \citet{inightetal23-1} as an AM\,CVn star based on its SDSS spectrum displaying a blue continuum with broad \Ion{He}{I} absorption lines and long-term variability in its ZTF light curve. \citet{kato23-1} detected a photometric 23\,min signal within the ZTF data, and suggested that this could be the orbital period of the system. We included this system in our survey to confirm the photometric period.

\begin{figure}
\includegraphics[angle=-90,width=\columnwidth]{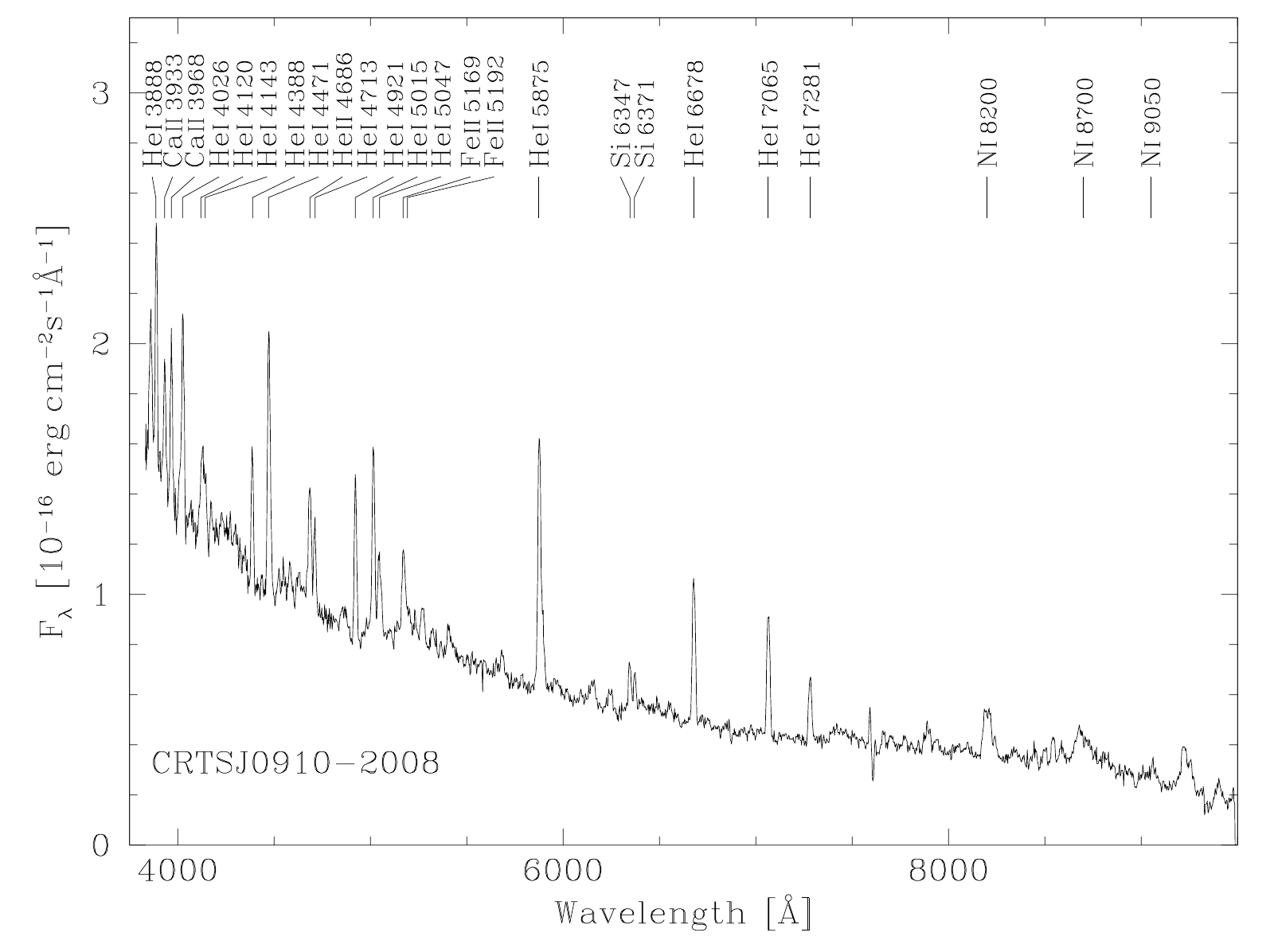} 
\includegraphics[angle=-90,width=0.98\columnwidth]{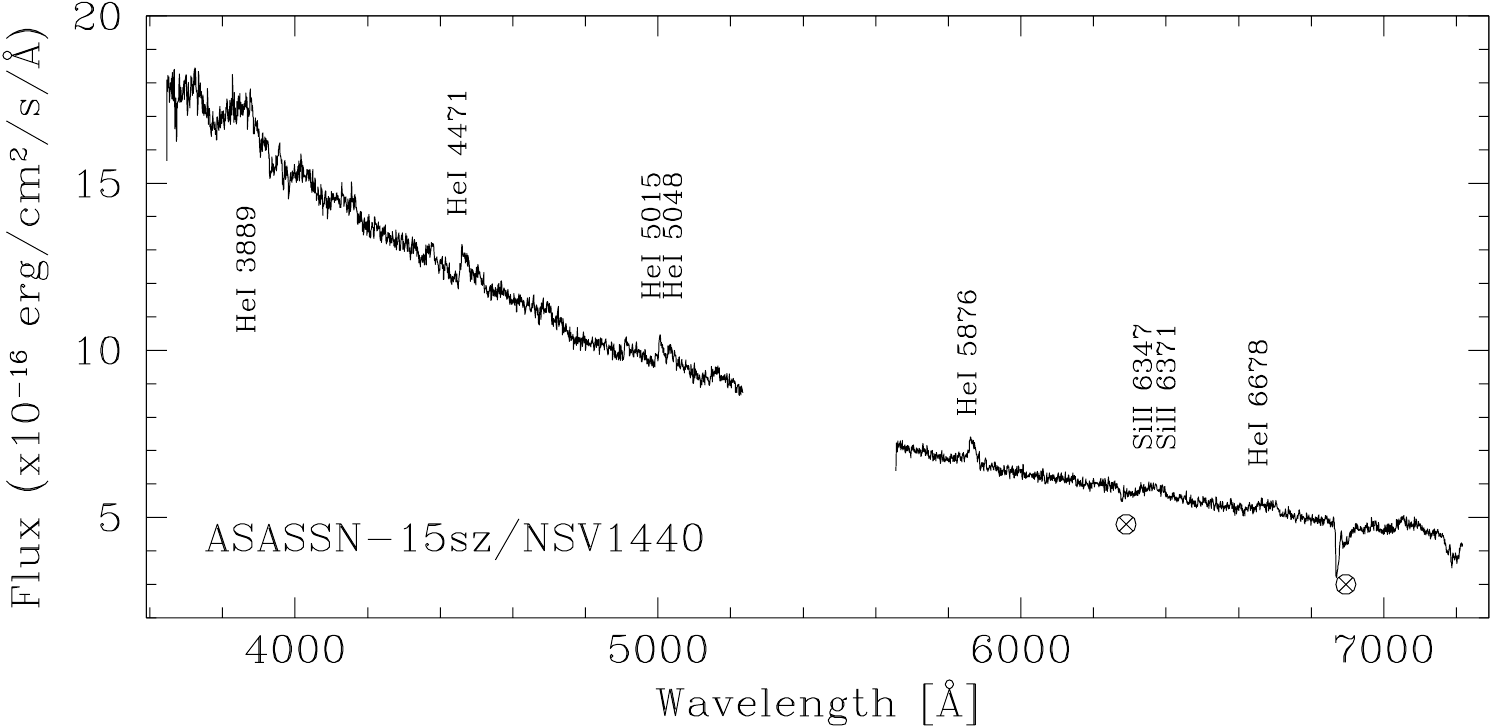}
\caption{\label{f-idspec} 
Identification spectra of CRTS\,J0910$-$2008 (top) and NSV\,1440 (bottom). The strongest emission lines are labelled, as well as oxygen absorption from the earth's atmosphere (crossed circle symbols).}
\end{figure}

\begin{figure}
\centerline{\includegraphics[angle=-90,width=\columnwidth]{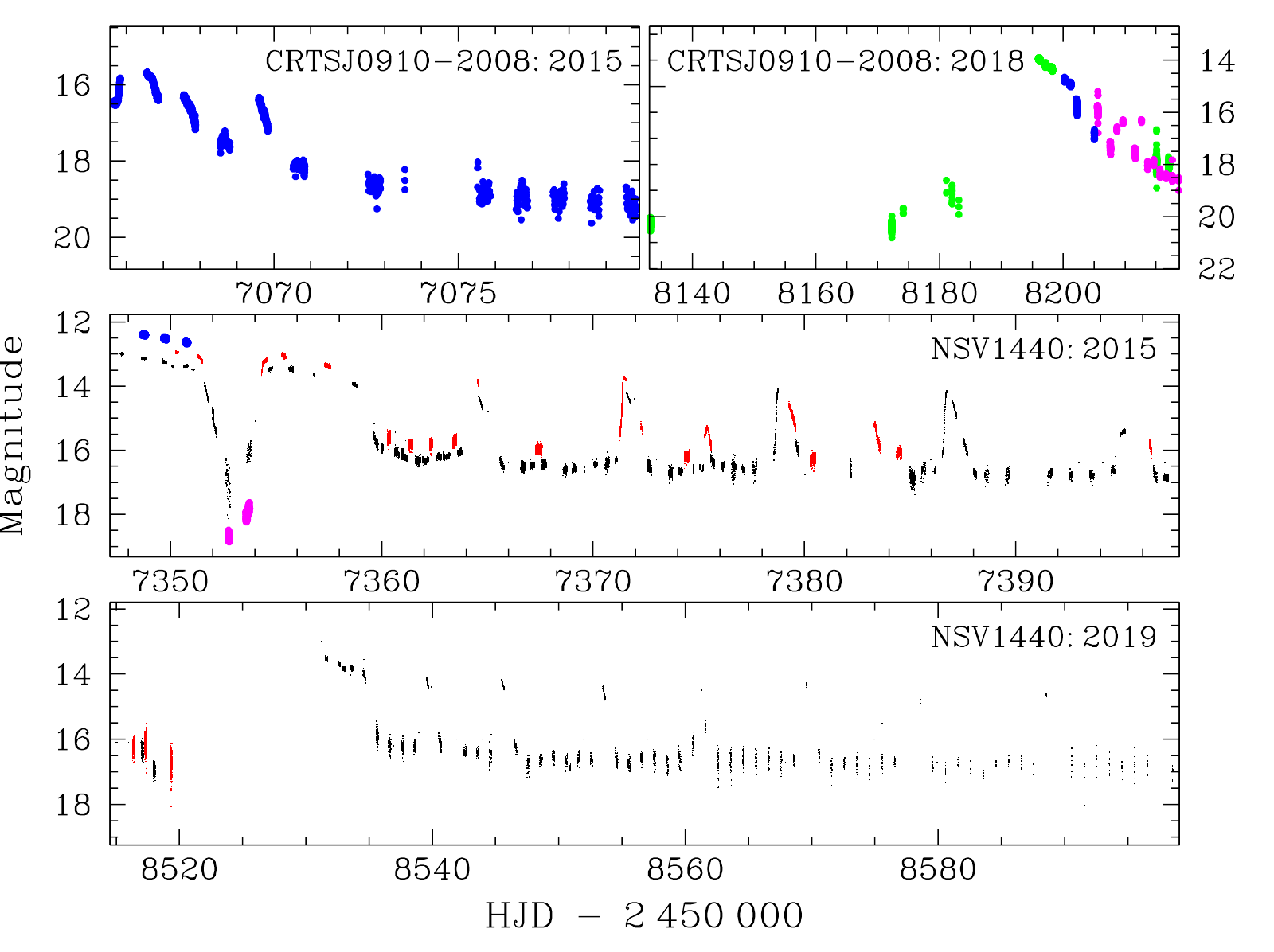}}
\caption{\label{f-outburst_mag} \textit{Top panels:} Our complete light curve data of the 2015 and 2018 outbursts of CRTS\,J0910$-$2008 obtained in $V$ (blue), $KG5$ (green) and Clear (magenta) filters. \textit{Bottom two panels:} The 2015 and 2019 AAVSO light curves of NSV\,1440 with $V$- and $R$-band data shown in black and red, respectively. Overlayed is our data in the $V$ (blue) and Clear filter photometry (magenta).}
\end{figure}

\section{Observations} \label{sec-obs}

\begin{figure*}
\centerline{
\includegraphics[width=0.45\textwidth]{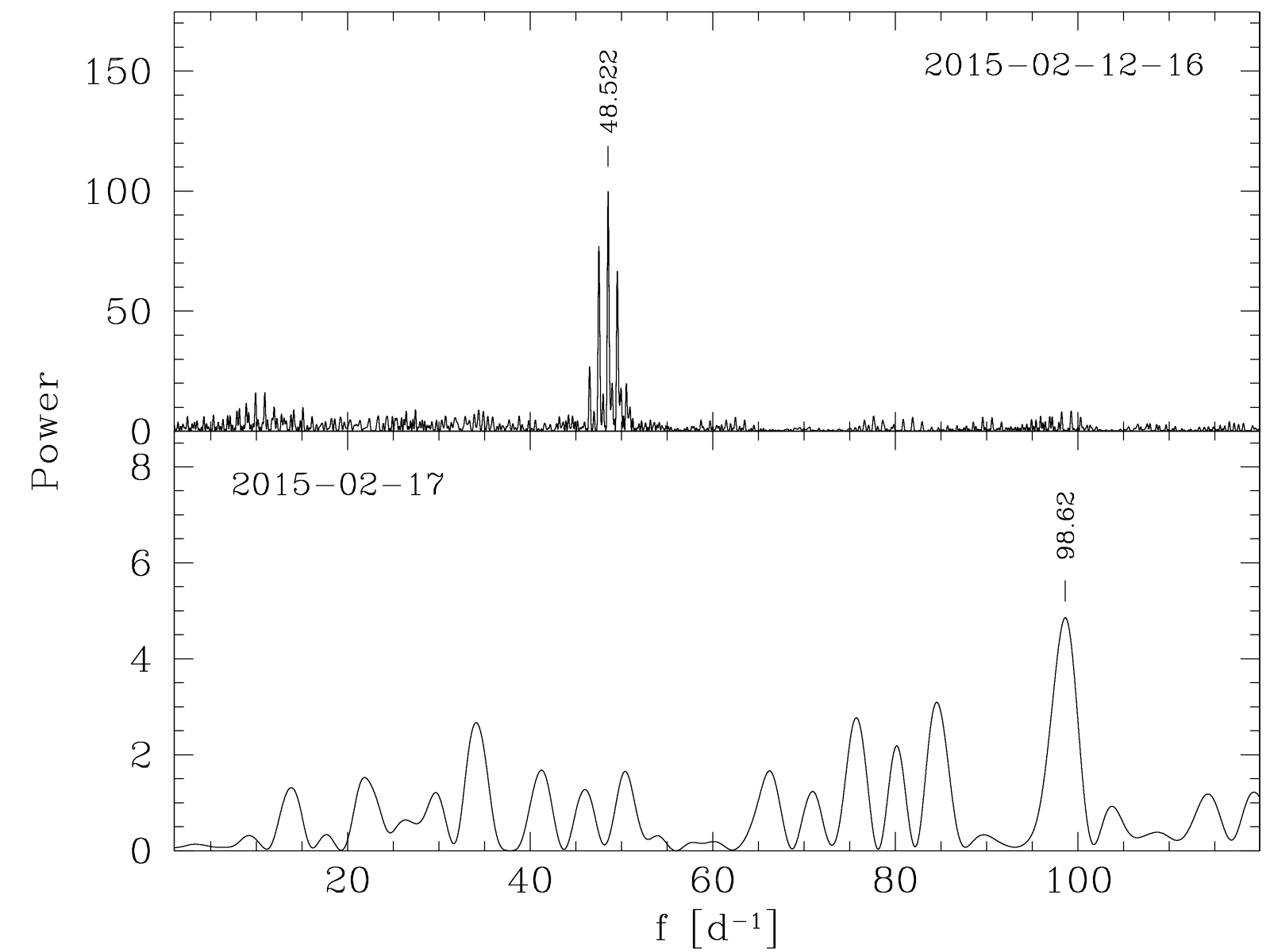}
\includegraphics[width=0.45\textwidth]{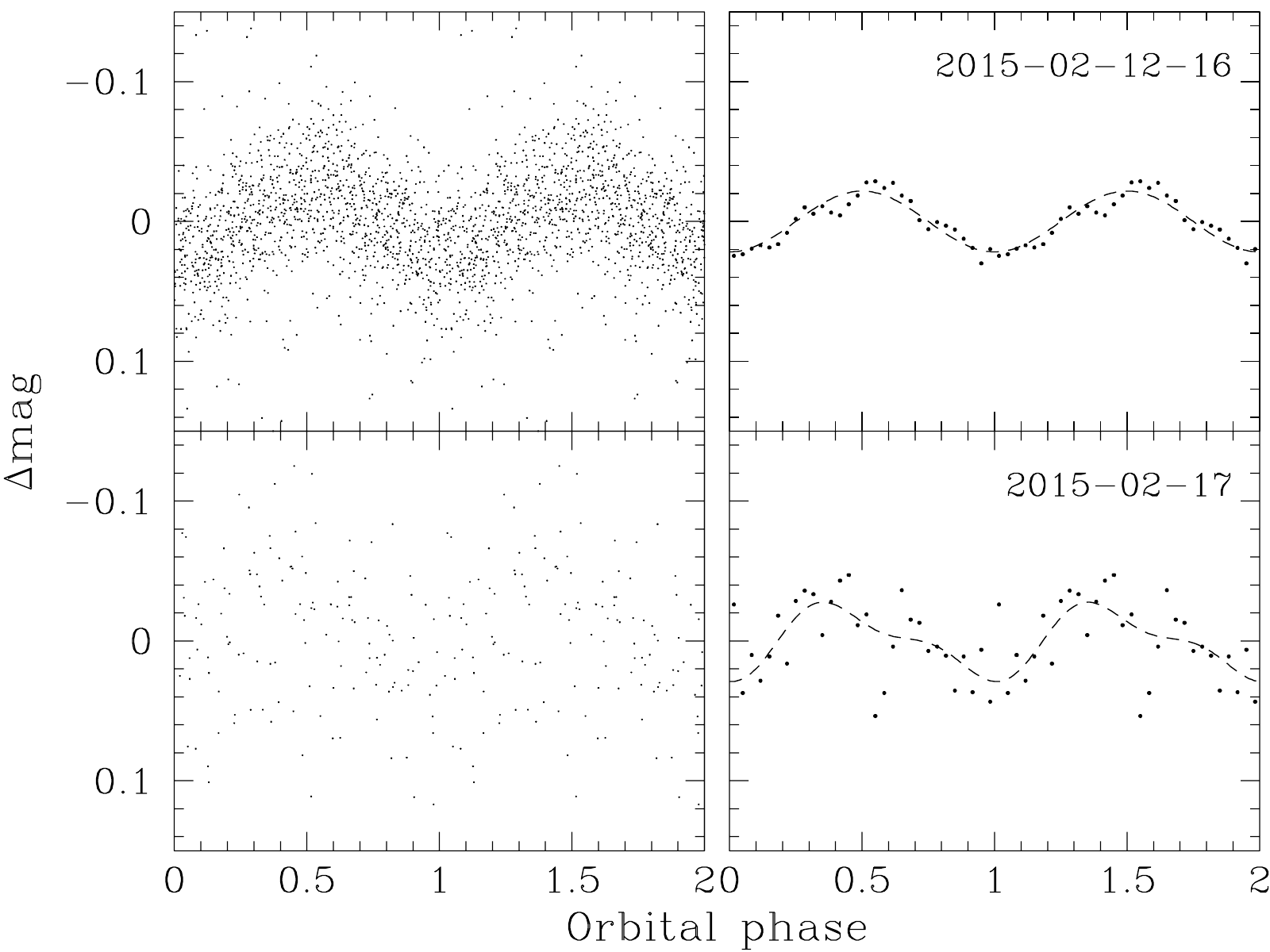}}
\centerline{
\includegraphics[angle=0,width=0.45\textwidth]{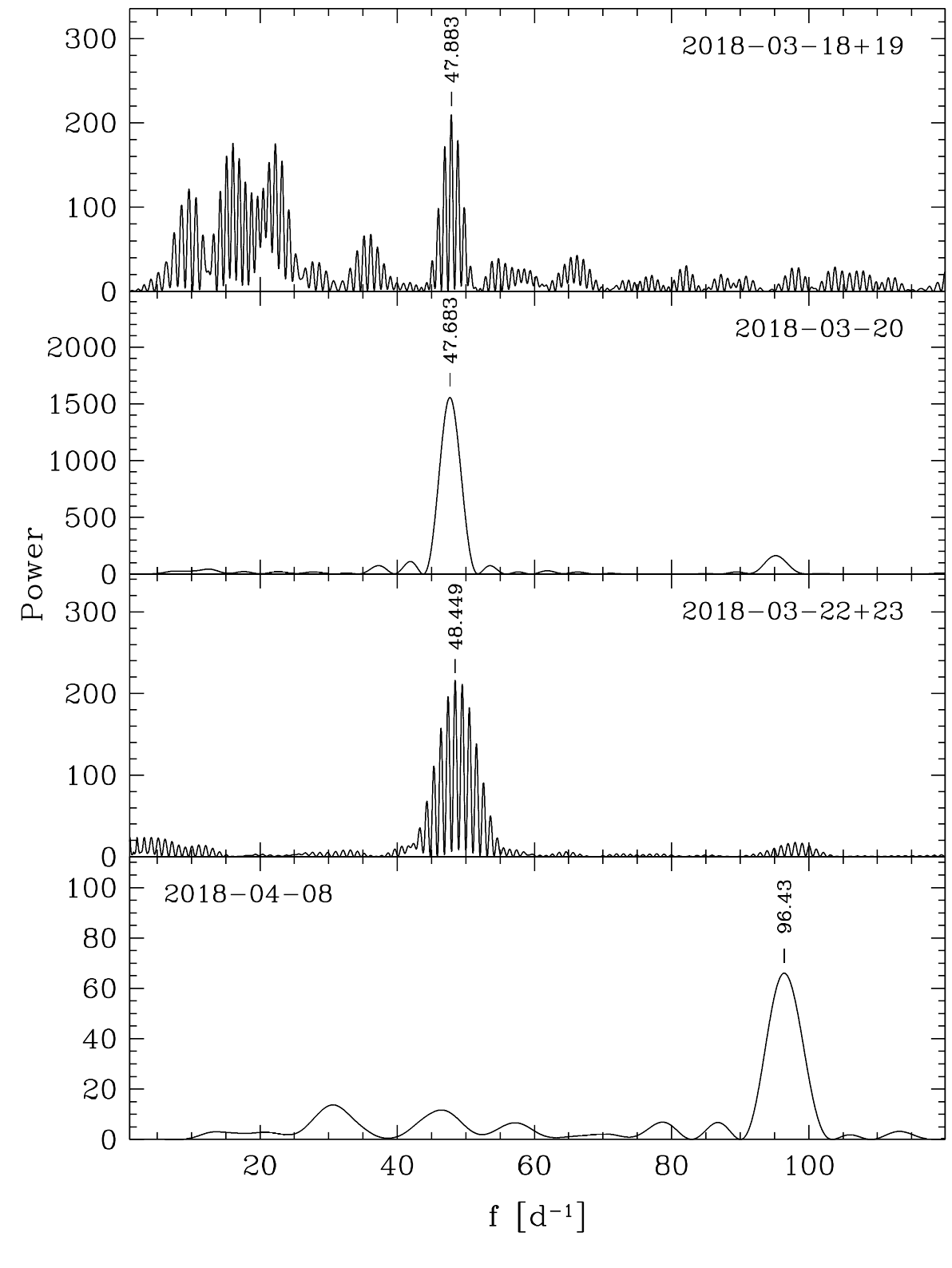}
\includegraphics[angle=0,width=0.45\textwidth]{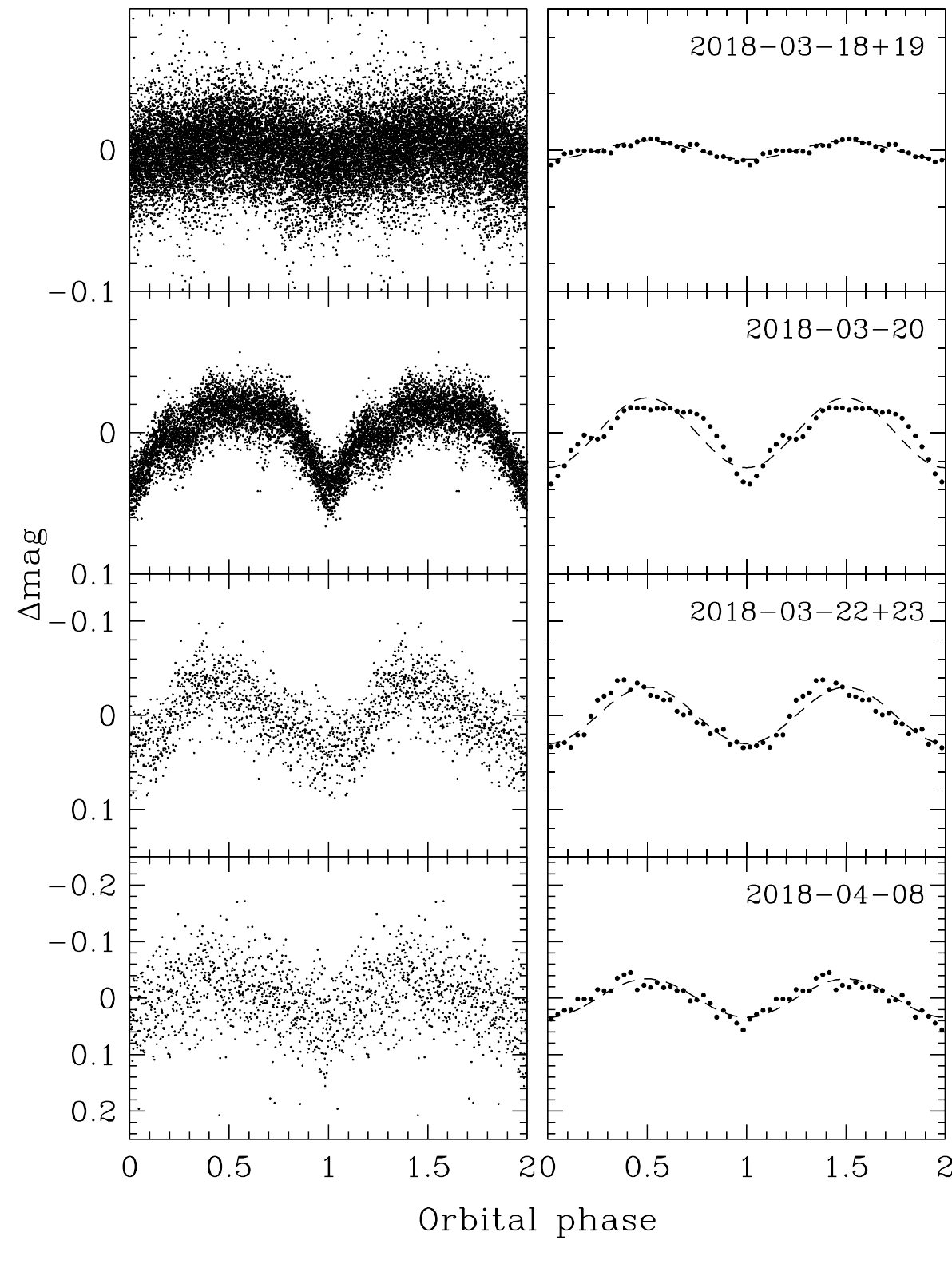}}
\caption{\label{f-scargle_css0910} 
Lomb Scargle periodograms (left panels) and phase-folded light curves (right two panels, actual data and binned into 30 phases) of CRTS\,J0910$-$2008 during the 2015 outburst (top two panels) and the 2018 outburst (bottom four panels). The dates or date ranges of the data used for the individual analyses are indicated near the top of each panel. Most of the data show a strong signal at $f\simeq48\,\id$, which we interpret as the superhump period of the system, $\Psh\simeq30$\,min. On two occasions (2015 February 17 and 2018 April 8) we detected a signal at $f\simeq97\,\id$ which we interpret as the second harmonic of the superhump.}
\end{figure*}
%

\subsection{Time series photometry}

Our follow-up photometry was carried out at the Cerro Tololo Inter-American Observatory in Chile, using the 0.6\,m Panchromatic Robotic Optical Monitoring and Polarimetry Telescopes (PROMPT8), at the Thai National Observatory (TNO) on Doi Inthanon, Thailand, using the 2.4-m TNT as well as the 0.5-m-TNO telescopes, and at the 0.4-m telescope of the Remote Observatory Atacama Desert (ROAD), San Pedro, Chile. PROMPT8 and the TNO are both operated by the National Astronomical Research Institute of Thailand (NARIT). ROAD is a privately owned telescope hosted at SPACEOBS\footnote{\href{http://www.spaceobs.com}{San Pedro de Atacama Celestial Explorations}}. A full log of the observations is presented in Tables\,\ref{t-obslog} -- \ref{t-obslog3}.

\subsubsection{CRTS\,J0910$-$2008}
CSS detected a $\approx3.4$\,mag outburst of CRTS\,J0910$-$2008 on 2015 February 11 and reported it to their transients webpage. We immediately started taking time series photometry on PROMPT8, observing the target during the period 2015 February 12--27 whenever the weather allowed, for approximately 4--8\,h per night. The PROMPT8 telescope is equipped with a $2$\,k $\times$ $2$\,k Apogee CCD with a pixel scale of $0.70\,\mathrm{arcsec\,pixel^{-1}}$, providing a $23.8\,\mathrm{arcmin}\times23.8\,\mathrm{arcmin}$ field of view. The images were taken in $V$ filter in unbinned mode. The raw images were bias-subtracted, dark current-subtracted, and flat-fielded using the SKYNET automated reduction pipeline. We extracted aperture photometry using {\sc sextractor} \citep{bertin+arnouts96-1} and converted the instrumental magnitudes to apparent magnitudes using the comparison stars listed in Table\,\ref{t-obslog}. 

Additional observations were carried out with the TNT from 2018 to 2024, the data were reduced using the ULTRACAM reduction pipeline, described in \citet{dhillonetal07-1}. We detected another outburst on 2018 March 18 at a brightness of $KG5=14.0$. A full log of the observations is presented in Table\,\ref{t-obslog}.

\subsubsection{NSV\,1440}
We used PROMPT8 to obtain time-series photometry of the system during 2015 November 22--27. The images were taken in $V$ filter on the first three nights while the object was bright ($V=12.4-12.6$), and filter-less on the next two nights when the object had dropped to a faint state at $\simeq18$\,mag. The data were reduced and extracted in the same manner as for CRTS\,J0910$-$2008 described above. 

Fig.\,\ref{f-surveylcs} shows that ASAS-SN detected several additional bright outbursts of this object. In this paper, we present our own observations of the 2015 outburst which were not included in \citet{isogaietal19-1}, as well as an analysis of the 2019 outburst, which was observed with the ROAD 0.4-m telescope for 68\,days. A log of the observations is given in Table~\ref{t-obslog2}.

\subsubsection{SDSS\,J1831+4202}
We obtained time-series photometry of SDSS\,J1831+4202 using ULTRASPEC \citep{dhillonetal14-1} at the TNT. All images were taken using the $KG5$ filter with $2\times2$ binning mode. The data were reduced using HiPERCAM data reduction pipeline\footnote{https://github.com/HiPERCAM/hipercam}. A superhump-like structure was detected on 2023 May 3  when the object was at $17.1$\,mag. The detailed log of the observations is given in Table~\ref{t-obslog3}.

\begin{figure*}
\centerline{
\includegraphics[width=\columnwidth]{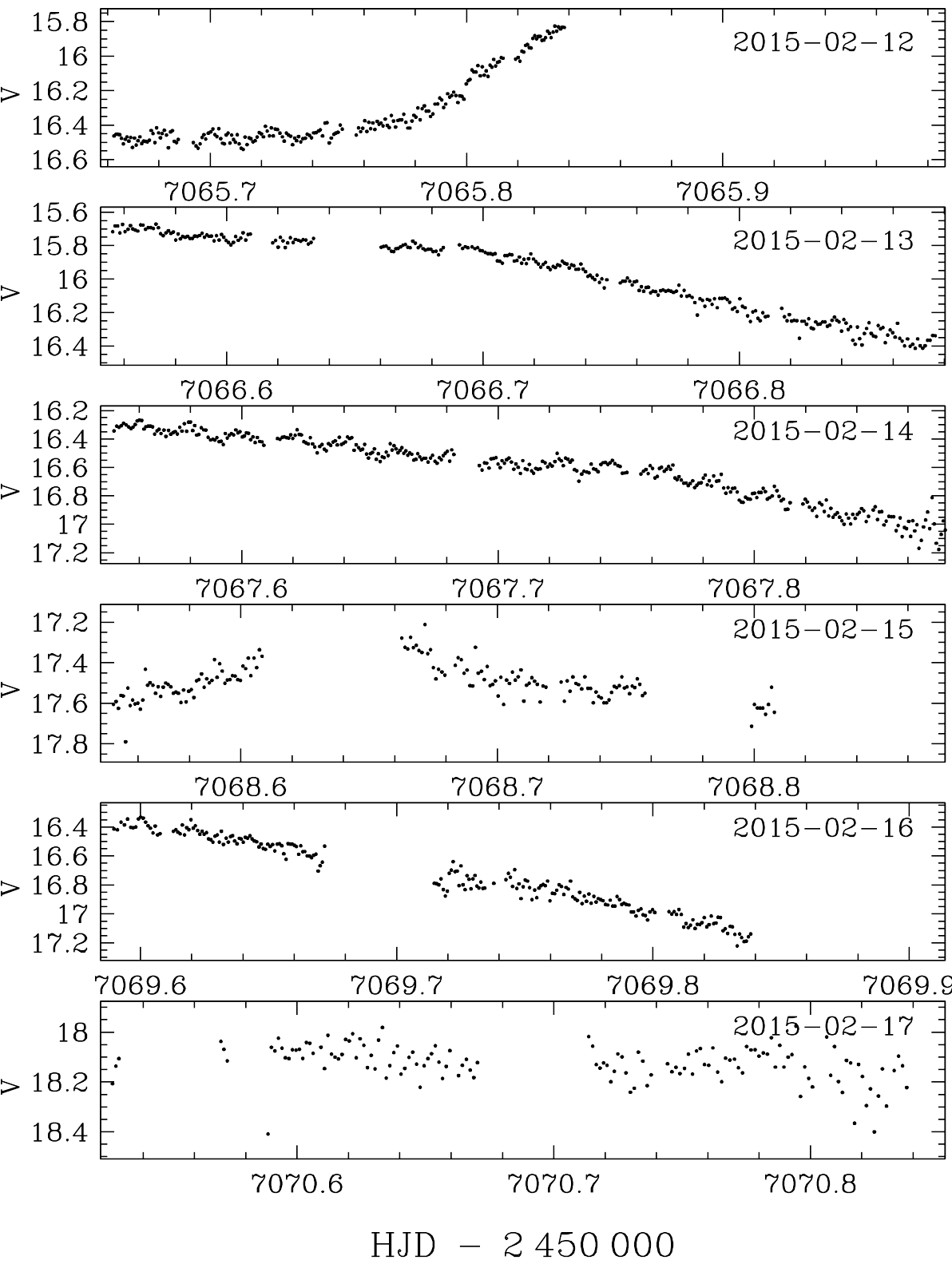}
\includegraphics[width=\columnwidth]{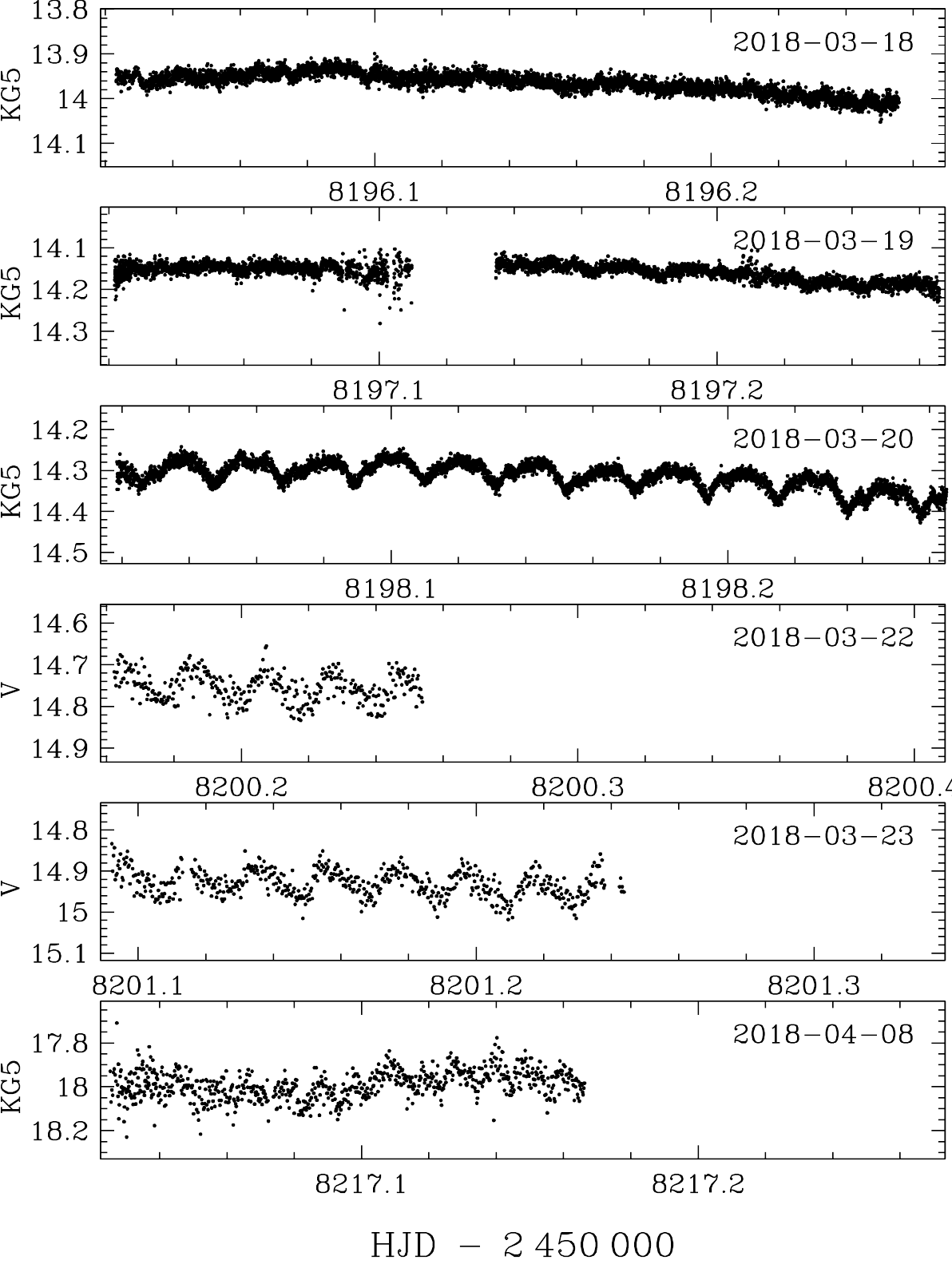}}
\caption{\label{f-lc_css0910} Light curves of CRTS\,J0910$-$2008 during the 2015 and 2018 outbursts used for superhump-period analysis. \textit{Left:} 2015 outburst. The light curves obtained on 12--14 February are dominated by a $\simeq15$\,min modulation superimposed on the declining night-to-night trend in the brightness of the system as it declines from the outburst. The system briefly re-brightened on February 16 (see Fig.\,\ref{f-surveylcs}) before fading into quiescence.  \textit{Right:} During the 2018 outburst we detected superhump variability throughout many subsequent nights.}
\end{figure*}

\subsection{Identification spectroscopy} \label{sec-idspec}

CRTS\,J0910$-$2008 and NSV\,1440 were identified as AM\,CVn candidates because of their short-period photometric variability (Section~\ref{sec-analysis}), and we obtained identification spectroscopy to confirm this classification.  

We obtained an optical spectrum of CRTS\,J0910$-$2008 at the European Southern Observatory's Very Large Telescope on 2015~May~5, using the FORS2 spectrograph. The spectrum shown in Fig.\,\ref{f-idspec} is the average of four 600\,s exposures, two taken using the GRIS\_300V+10 grism and two using the GRIS\_300I+11 grism. The combined spectrum covers the range $3830 - 10110$\,\AA\ at $12$\,\AA\ resolution. Conditions were clear at the time of the observation, but morning twilight was too cloudy to take a standard star spectrum, so we derived an approximate flux calibration using an archival observation of the standard star LTT3864, taken with the same instrumental setup. 

For NSV\,1440, we obtained low-resolution optical spectra using a 930 line$\:$mm$^{-1}$ grating with the Goodman spectrograph on the 4.1-m SOAR telescope \citep{clemensetal04-1} on the night of 2016 January 8. Using two different grating and camera angles, we obtained setups that cover the wavelength range $3600-5200$\,\AA\ ($4\times300$\,s exposures) and $5640-7200$\,\AA\ ($4\times300$\,s exposures), with a dispersion of roughly 0.84~\AA~pixel$^{-1}$. We used a 3.2\,arcsec slit, so our spectral resolution is seeing limited, roughly 4\,\AA\ in the mean 1.3\,arcsec seeing during our observations. We flux calibrated our spectra using the spectro-photometric standard GD\,108.

The reduction of all spectra and calibration frames were done using standard {\sc starlink} routines \citep{currieetal14-1} and then the spectra were optimally extracted using the {\sc starlink} package {\sc pamela} \citep{marsh89-1}. The wavelength and flux calibration, as well as a heliocentric correction to the time stamps, were applied using {\sc molly}\footnote{{\sc molly} was written by T.\,R.~Marsh and is available from  http://www.warwick.ac.uk/go/trmarsh/software/}. 

The spectrum of CRTS\,J0910$-$2008 is dominated by strong helium emission, along with emission lines from He, Ca, Si, N and Fe, as shown in Fig.\,\ref{f-idspec}. The feature at 7600\,\AA\, is residual telluric absorption. No lines from hydrogen are detected in the spectrum, consistent with those of other AM\,CVn stars.   

The emission features are weaker in NSV\,1440, but we detect lines of He, Si and Ca in this spectrum as well (Fig.\,\ref{f-idspec}). No feature from H is detected. This confirms the classification of NSV\,1440 as an AM\,CVn star.

\begin{figure}
\begin{center}
\includegraphics[width=\columnwidth]{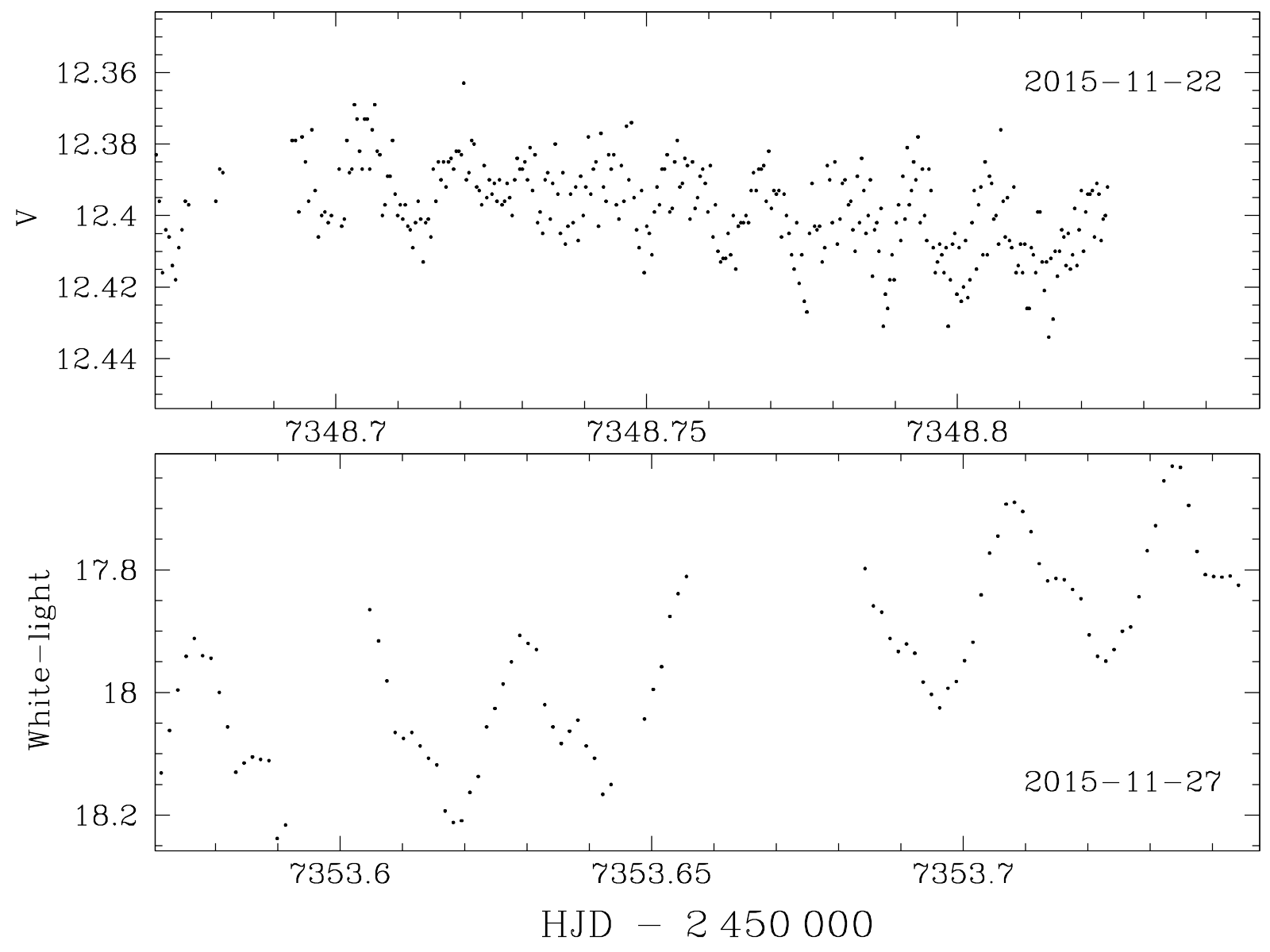}
\includegraphics[width=\columnwidth]{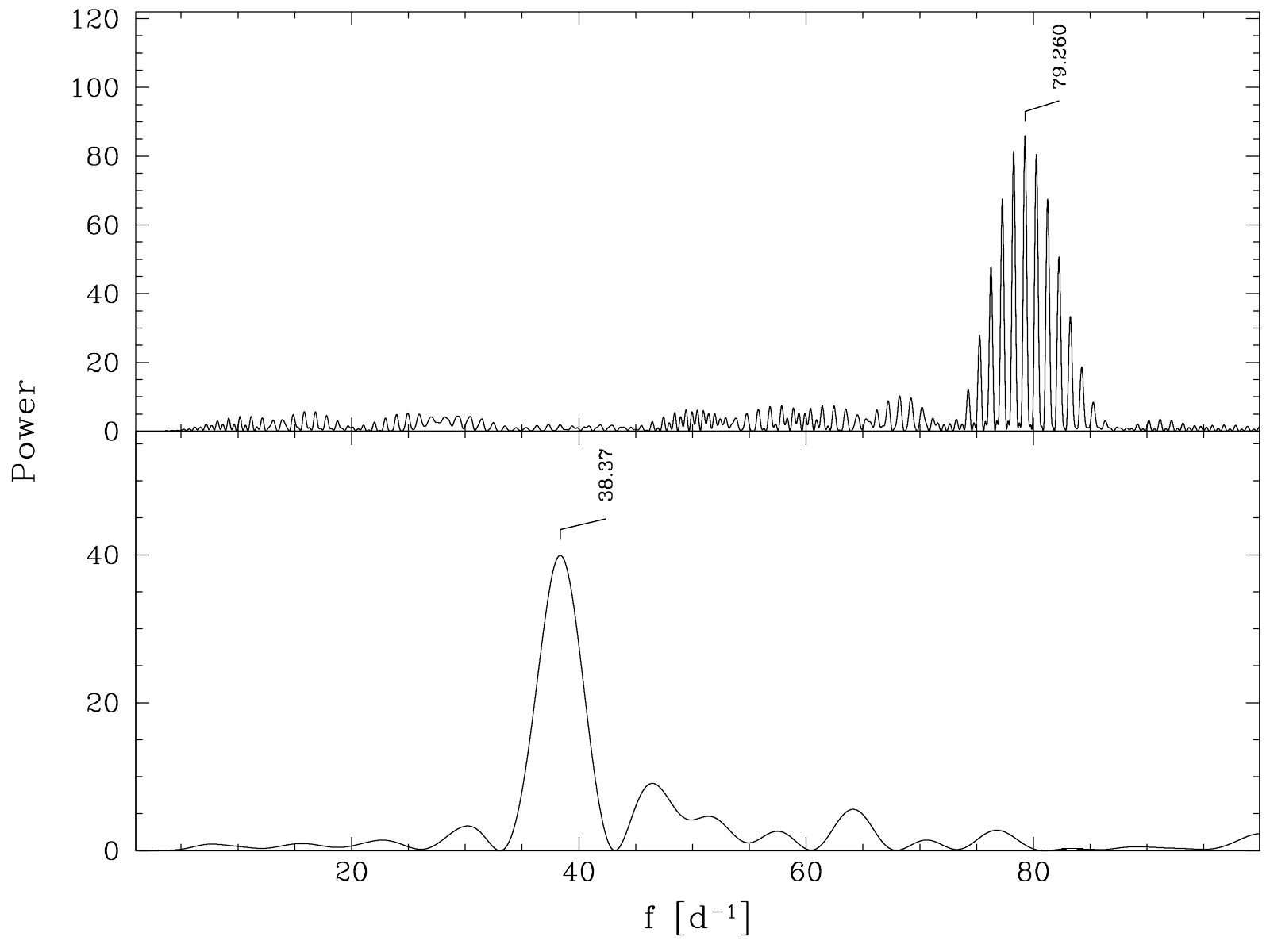}
\includegraphics[width=\columnwidth]{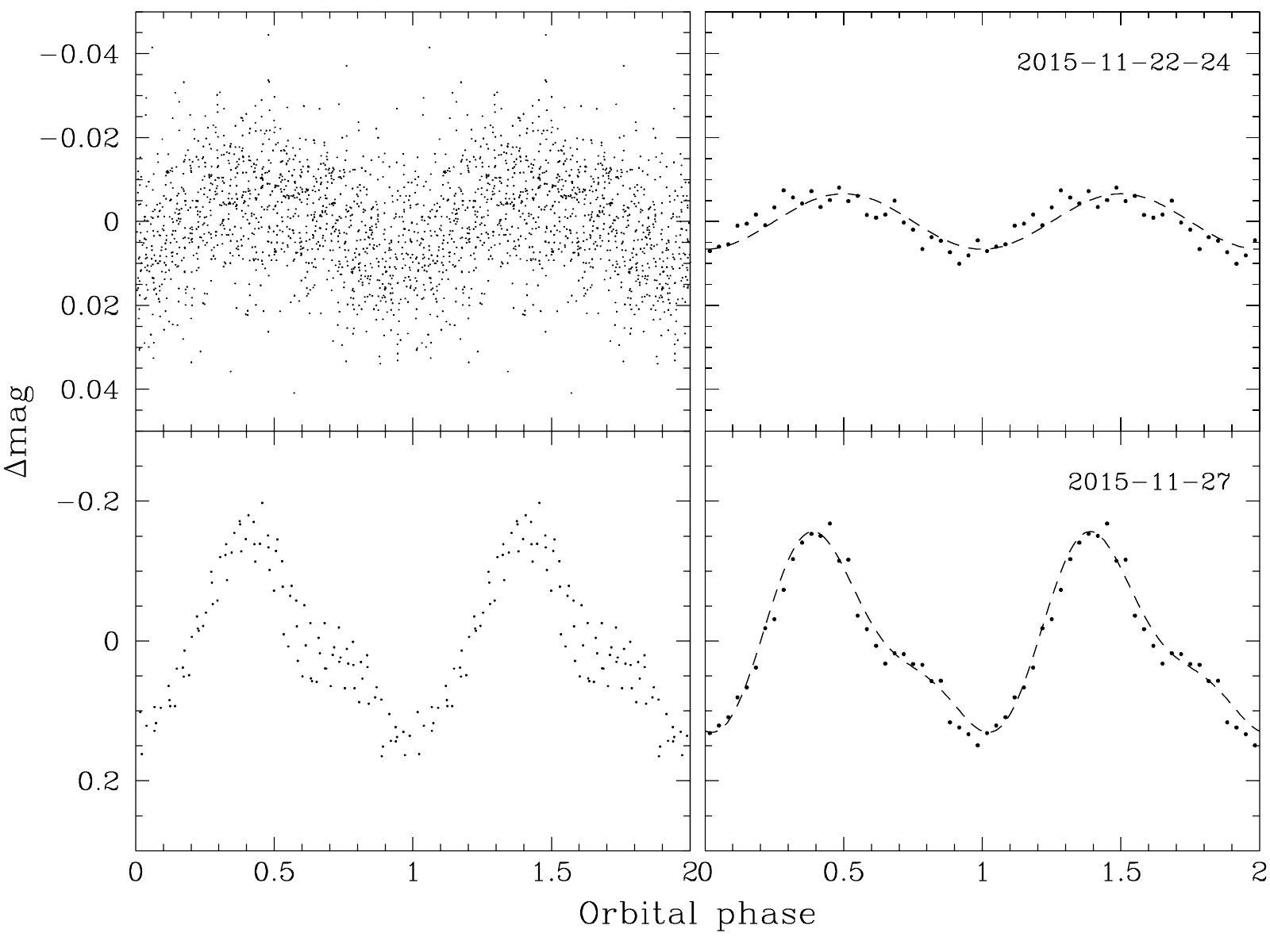}
\end{center}
\caption{\label{f-nsv1440}  
\textit{Top panels:} The light curve of NSV1440 obtained on 2015~November~22 when the object was brightest exhibits clear short time-scale variations. By November 27 the object dropped in brightness by $\sim6$\,mag displaying larger-amplitude superhump structures in the light curve. Note that the scales on the horizontal axes are the same, so the periods of the variations are directly comparable.
\textit{Middle panels:} Scargle periodograms computed from the combined data obtained on 2015 November 22--24 (peak frequency $f=79.260$\,\id), and November 27 (peak frequency $f=38.37$\,\id). 
\textit{Bottom panels:} The corresponding phase-folded light curves (left: individual observations; right: data binned into 30 phases) using the strongest signal detected in each of the two data sets. The superhump structure is clearly seen in the light curve.}
\end{figure}

\section{Analysis}\label{sec-analysis}
\subsection{CRTS\,J0910$-$2008}
We observed CRTS\,J0910$-$2008 in outburst twice. Our observations covered a period of 13 days in 2015 February and 27 days in 2018 March. The complete light curves for the two outbursts are shown in Fig.\,\ref{f-outburst_mag}.

\subsubsection{2015 outburst}

We obtained 4\,h of time-series photometry on February 12, which showed that the object was on the rise to outburst, brightening by $\simeq0.4$\,mag towards the end of the night (Fig.\,\ref{f-lc_css0910}). A short-time scale variation of $\simeq30$\,min was also visible, suggesting the object might be a system below the conventional period minimum of cataclysmic variables. In order to confirm the nature of the object, we continued our observations for as long as possible. The light curves on February 13--14 covered $\simeq7.5$\,h each night, and both presented the same trend of declining brightness ($\sim0.7-0.9$\,mag) plus a consistent signal of short-time scale modulation of $\simeq30$\,min as seen on the first night (Fig.\,\ref{f-lc_css0910}). On February 15, the object dropped in brightness by nearly one magnitude, but it re-brightened again on the next night, with a light curve very similar to those of February 13--14. Overall the light curves of CRTS\,J0910$-$2008 on February 12--16 can be described as a $\simeq0.1$\,mag, $\simeq30$\,min quasi-periodic modulation with a `superhump-like' structure, superimposed on a trend of $\sim0.5-1.0$\,mag night-to-night variation. The superhump structure, however, disappeared on February~17 when the object dropped to $\simeq18.1$\,mag.  A short-time scale variation of $\simeq15$\,min can be noticed in the light curve on this night. For February 19--27, the object remained in a faint state of $\geq18$\,mag, which reaches telescope limit. We obtained five images of $100$\,s each every 20\,min to monitor the evolution of the outburst, however these light curves were too short to confirm $\simeq15$\,min signal. The two-week light curve shown in the top panel of Fig.\,\ref{f-outburst_mag} shows that the object returned to its quiescent state within a week after the outburst was detected by CSS. 

The superhump periods measured on the individual nights from February 12--16 were very similar, so we combined the data from these five nights. We detrended the light curves for the overall variations in the average brightness of the system and then computed a \citet{scargle82-1} periodogram using the \texttt{MIDAS/TSA} context. The strongest signal is found at $f=48.522\pm0.007$\,\id\ where the error is determined from fitting a sine wave to the data. We identified this signal as a superhump period $\Psh=29.677\pm0.004$\,min (Fig.\,\ref{f-scargle_css0910}, top left panel). The phase folded light curve (Fig.\,\ref{f-scargle_css0910}, right panels) has a quasi-sinusoidal morphology also seen in the individual light curves. In addition, we also computed a Scargle periodogram of the February 17 light curve, after detrending the low frequency variation (Fig.\,\ref{f-scargle_css0910}, second-left panel). It reveals a weak signal at $f=98.62\pm0.65$\,\id\ or $P=14.6\pm0.1$\,min. This is nearly double the peak frequency of the nights before, likely a harmonic of the superhump signal.

\subsubsection{2018 outburst}
In 2018 January to March, we obtained TNT time-series photometry of CRTS\,J0910$-$2008 in its quiescent state ($\sim19.0-20.4$\,mag) to probe for variability (see Table~\ref{t-obslog}). Overall the light curves of CRTS\,J0910$-$2008 during quiscence display small variations of $\sim0.4-0.6$\,mag. A Scargle periodogram of the $\simeq6$\,h quiescent data suggests a possible periodicity at $\sim32.9$\,min, but there are many alias signals so additional photometry is necessary to confirm this periodicity. On March 18, we detected CRTS\,J0910$-$2008 in outburst again. It had a mean magnitude of $\sim14.0$, which is brighter than the mean magnitudes of $\sim15.0-16.0$ of the previous outbursts detected by CSS and SSS. We followed the outburst over a total of 21 nights with a variety of telescopes, as weather and telescope access allowed (Table\,\ref{t-obslog}). We show examples of the superhump variability detected during the period of March 18 and April 8 in Fig.\,\ref{f-lc_css0910}.

A weak short-period variability of $\simeq30$\,min appeared in the March 18 and 19 data and it grew to a clearly visible superhump signal in the March 20 light curve. This superhump signal dominated the light curves of March 22--23 as the object brightness declined to $V=14.8-14.9$. On March 24, the object's brightness dropped by one magnitude and the superhump was less obvious in our $\simeq1$\,h long light curve, we therefore did not include that night in the analysis. In April, we detected the system at $\simeq18.0$\,mag. The light curve on April 8 reveals a short-timescale variation of $\simeq15$\,min as found in the 2015 outburst. This short-timescale brightness variation is shown in the bottom panel of Fig.\,\ref{f-lc_css0910}.  

In order to study the evolution of the superhump during the 2018 outburst, we created Scargle periodograms for several subsets of data, depending on morphology of the light curves (Fig.\,\ref{f-scargle_css0910}). As before, we removed the low-frequency trends in the light curves before calculating the periodograms. The March 18 and 19 light curves display low amplitude, short-period variations and their Scargle periodogram confirms this, with a peak at $f=47.883\pm0.015$\,\id, or $30.073\pm0.009$\,min. By March 20, the amplitude of the variations had grown to 0.1\,mag and the Scargle periodogram indicated a slightly longer period of $P=30.199\pm0.016$\,min, or $f=47.683\pm0.025$\,\id. The combined light curves from March 22 and 23 gave a peak in the Scargle periodogram at $f=48.449\pm0.012$\,\id, or $\Psh=29.722\pm0.007$\,min consistent with the superhump period from the 2015 outburst. We take the average of the two values as the superhump period of CRTS\,J0910$-$2008, $\Psh=29.700\pm0.004$ (see Table\,\ref{t-Psh} for a summary). Finally, the Scargle periodogram of the April 8 data has a peak at $f=96.43\pm0.30$\,\id, or $P=14.93\pm0.05$\,min.

\subsection{NSV\,1440}

\subsubsection{2015 outburst}
We obtained time-series photometry of NSV\,1440 during 2015 November 22--27. The full light curve is shown in Fig.\,\ref{f-outburst_mag}. On November 22--24, the object was in very bright state, $V=12.4-12.6$, and then faded rapidly to 18.7\,mag on November 26, but rose again in brightness the next night. Poor weather on November 25 prevented us from taking any observations on that night. 

Our observations on November 22--24 revealed small amplitude variations ($\simeq0.02$\,mag, Fig.\,\ref{f-nsv1440}, top panel) with a period of $18.168\pm0.003$\,min ($f=79.260\pm0.014$\,\id) for the combined data set (Fig.\,\ref{f-nsv1440}, middle panel). Treating the nights individually yields periods close to this value, that agree with each other within the error bars. We interpret this signal as the harmonic of the superhump period, i.e. $36.336\pm0.003$\,min. This value is almost identical to the early superhump period reported by \citet{isogaietal19-1}.

The light curve on November 27 reveals a prominent superhump structure while the system was brightening overall by $\simeq0.3$\,mag (Fig.\,\ref{f-nsv1440}, second-top panel). The Lomb Scargle periodogram from these data shows a strong signal at $f=38.37\pm0.13$\,\id (Fig.\,\ref{f-nsv1440}, middle panel) which we define as the superhump period $\Psh=37.53\pm0.13$\,min, longer than the period of the ``stage A'' superhumps (36.98\,min) reported by \citet{isogaietal19-1}. The corresponding phase-folded light curves of these two frequencies are illustrated in Fig.\,\ref{f-nsv1440} (bottom panels).

\subsubsection{2019 outburst}
We started time-series observations immediately following the detection of another bright outburst of NSV\,1440 on 2019 February 16 (Stubbings 2019, vsnet-alert 23005). The system stayed bright ($V<14$) for four nights before fading back to $V\simeq16$. Over the course of a further 64 nights, we observed six rebrightenings, as well as one small ($\simeq1$\,mag) outburst (Fig.\,\ref{f-outburst_mag}), similar to the behaviour seen in the 2015 and 2017 outbursts \citep{isogaietal19-1}. A periodogram computed from the observations obtained during the first four days of the outburst shows a strong signal at $\simeq39.46\,\id$, corresponding to $P=36.525\pm0.017$\,min, similar to that of the ``stage B'' superhumps reported by \citet{isogaietal19-1}.

\begin{figure}
\centerline{
\includegraphics[width=\columnwidth]{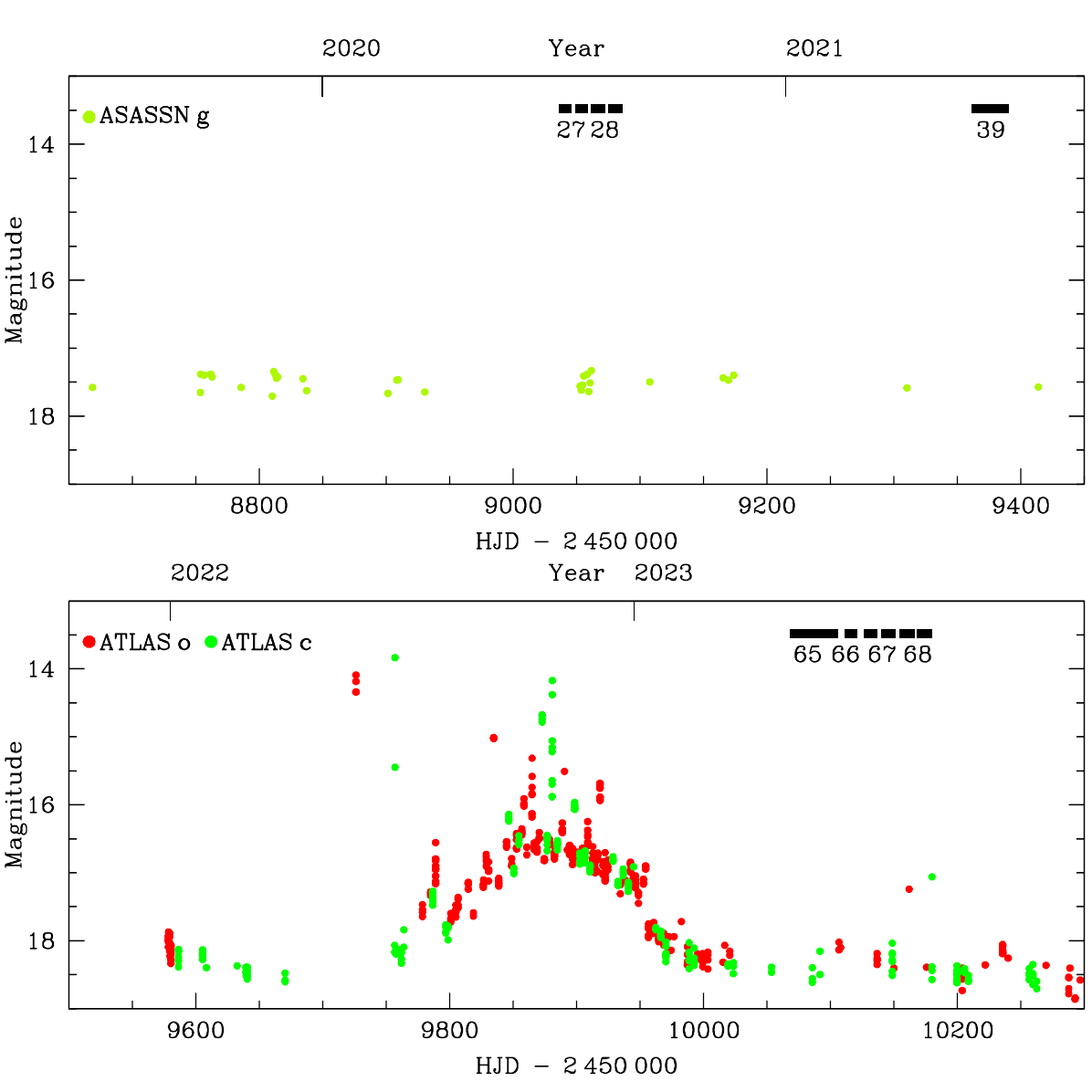}}
\caption{\label{f-atlas_tess_nsv1440} \textit{TESS} observed NSV\,1440 during quiescence in sectors 27, 28, and 39 (top panel, ASAS-SN $g$-band data in green). Late in 2022, the ATLAS cyan ($c$) and orange ($o$) bandpass light curves captured a bright state of NSV\,1440 lasting $\simeq200$\,d. Throughout this high state, normal outbursts continued which indicates that the disc, or sufficiently large parts of it, were still undergoing thermal disc instabilities. \textit{TESS} observed the system after its high state during sectors 65--68.}
\end{figure}

\begin{figure}
\centerline{
\includegraphics[width=\columnwidth]{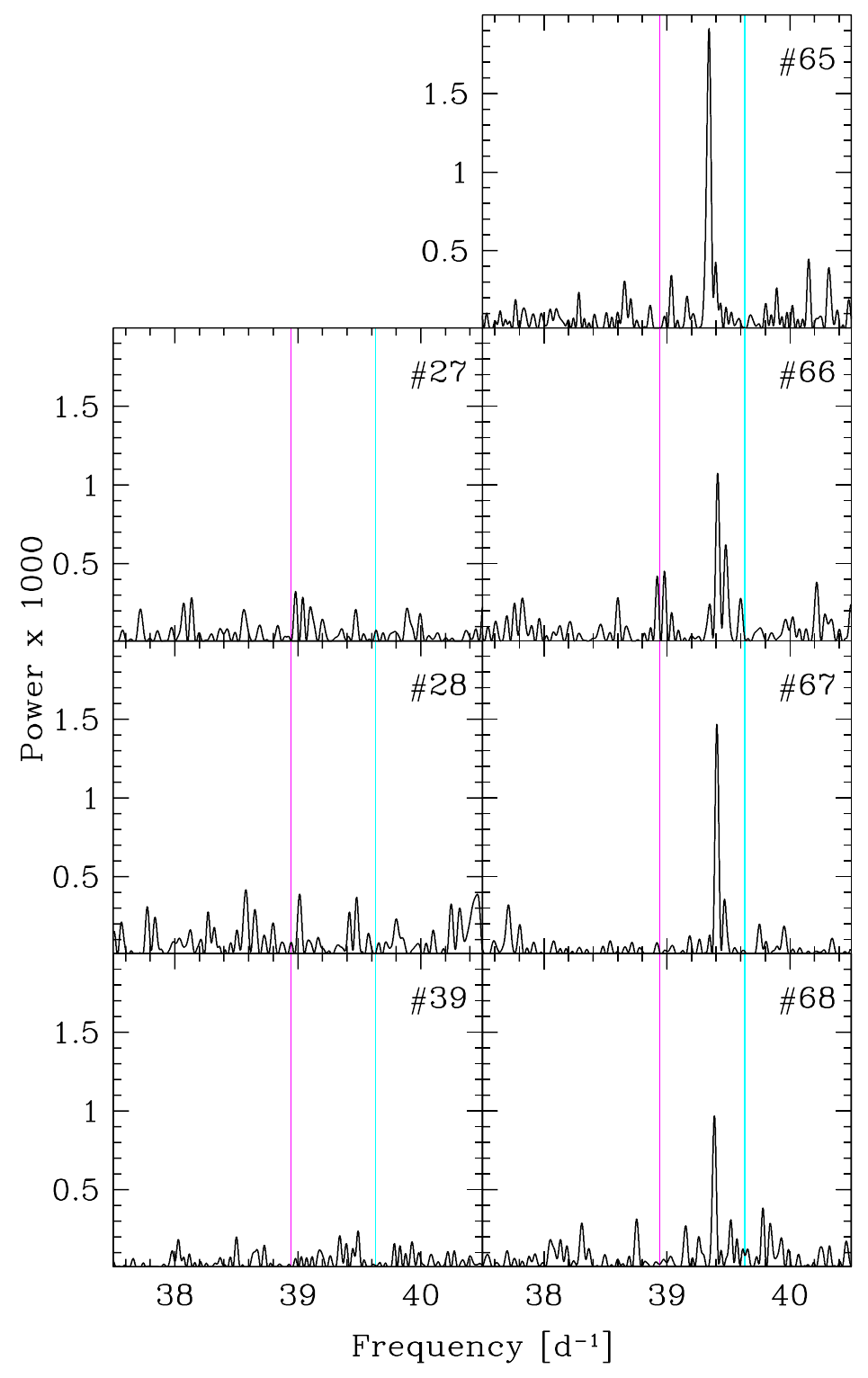}}
\caption{\label{f-tess_nsv1440_power} Power spectra computed from the seven sectors of \textit{TESS} observations of NSV\,1440. The magenta and cyan lines correspond to the early superhumps with $P=36.335\pm0.071$\,min) and the growing ``stage A'' superhumps with $P=36.978\pm 0.029$\,min of \citet{isogaietal19-1}, who interpreted the early superhump signals as the orbital period. Each \textit{TESS} light curve was normalised to a mean flux of unity prior to the analysis. Sectors 27, 28, and 39 were obtained in deep quiescence, and do not contain any significant signal. Sectors 65--68 were obtained $\simeq100-200$\,d after a ``high state'' and independently show a strong signal at $\simeq39.40\,\mathrm{d^{-1}}$, or $P=36.56$\,min. Given the stability of the period detected in the \textit{TESS} data, we argue that this signal corresponds to the orbital period of NSV\,1440.
}
\end{figure}

\subsubsection{\textit{TESS} observations and the 2022 ``high state''}
The \textit{TESS} mission observed NSV\,1440 in short-cadence mode (120\,s exposures) during sectors 27, 28, 39, and 65--68. The first three sectors were obtained during quiescence, and the last four sectors were obtained following a ``high state'' where NSV\,1440 was about two magnitudes brightener than its typical quiescent level (Fig.\,\ref{f-atlas_tess_nsv1440}). During that state, NSV\,1440 exhibited multiple short outbursts with amplitudes of $\simeq2$\,mag above the baseline brightness. 

Extremely long brightening events have been observed in a number of other AM\,CVn stars including SDSS\,J080710.33+485259.6 ($\Porb\simeq53$\,min \citealt{riverasandovaletal20-1}), SDSS\,J113732.34+405458.0 ($\Porb\simeq60$\,min, \citealt{riverasandovaletal21-1}) and ASASSN-21au ($\Porb\simeq58$\,min, \citealt{riverasandovaletal22-1}). All three systems have much longer periods than NSV\,1440, and the detailed morphology of their brightening events differs from that seen in NSV\,1440, having a slow rise followed by a fairly sudden fading, and a re-brightening event in the case of ASASSN-21au. The fact that NSV\,1440 exhibits normal outbursts throughout the bright state indicates that at least large parts of the disc continue to undergo thermal instabilities. The nature of the brightenings has been discussed in the context of episodes of enhanced mass transfer as well as thermal instabilities within the disc (\citealt{riverasandovaletal22-1}, and \citet{kato+stubbings23-1} for the case of NSV\,1440). However, no consensus has yet been reached on the exact physical origin of these events. 

We computed discrete Fourier transforms for all \textit{TESS} observations, finding no significant signal in the first three sectors, but detecting a strong signal at a mean value of $36.56\pm0.03$\,min in the last four sectors. Given the stability of this period throughout nearly four months of \textit{TESS} observations, we argue that this signal very likely represents the orbital period of the system, i.e. slightly longer than the value of $36.335\pm0.071$\,min adopted by \citet{isogaietal19-1}.

\subsection{SDSS\,J1831+4202}
\begin{figure}
\begin{center}
\includegraphics[width=\columnwidth]{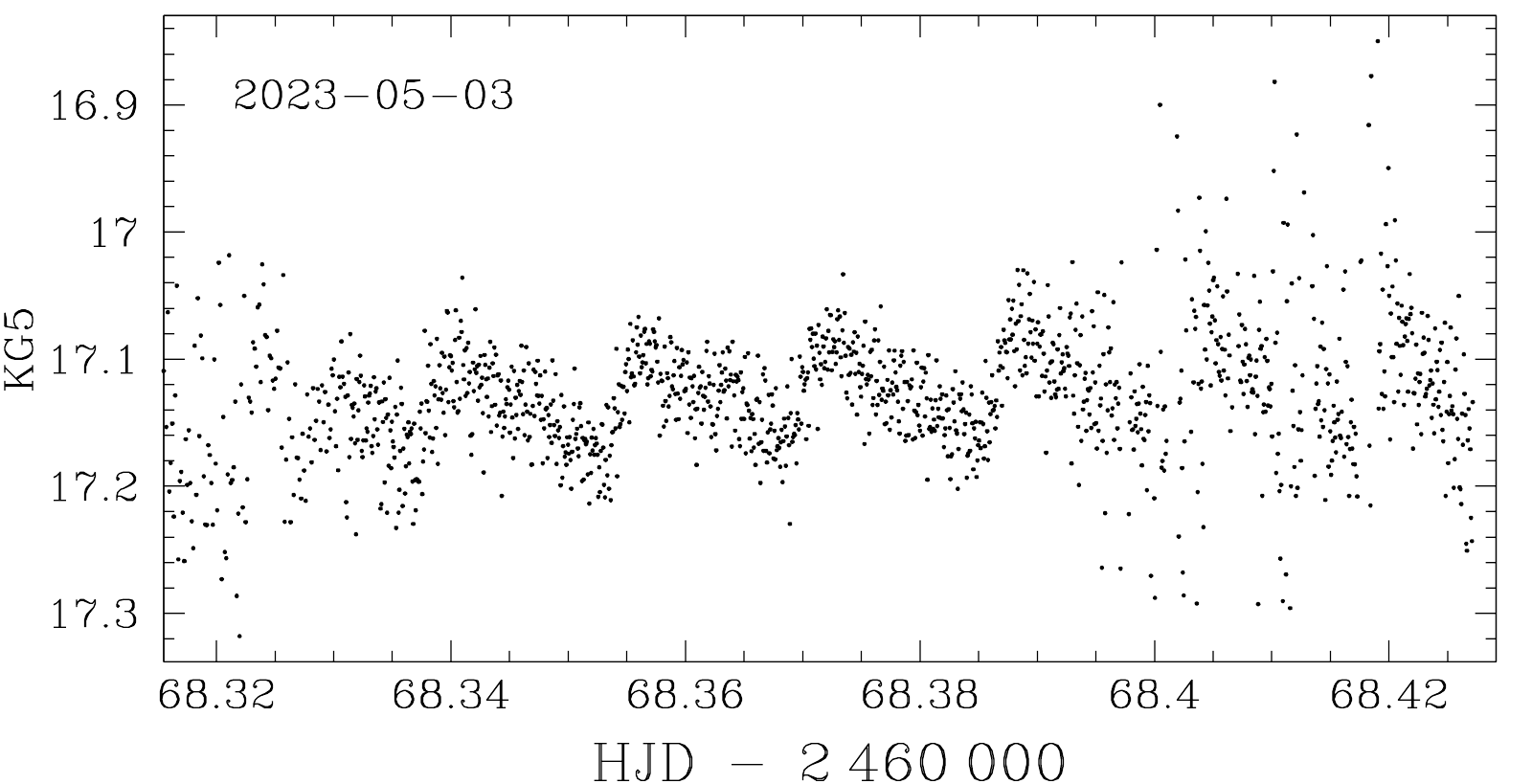}
\includegraphics[width=\columnwidth]{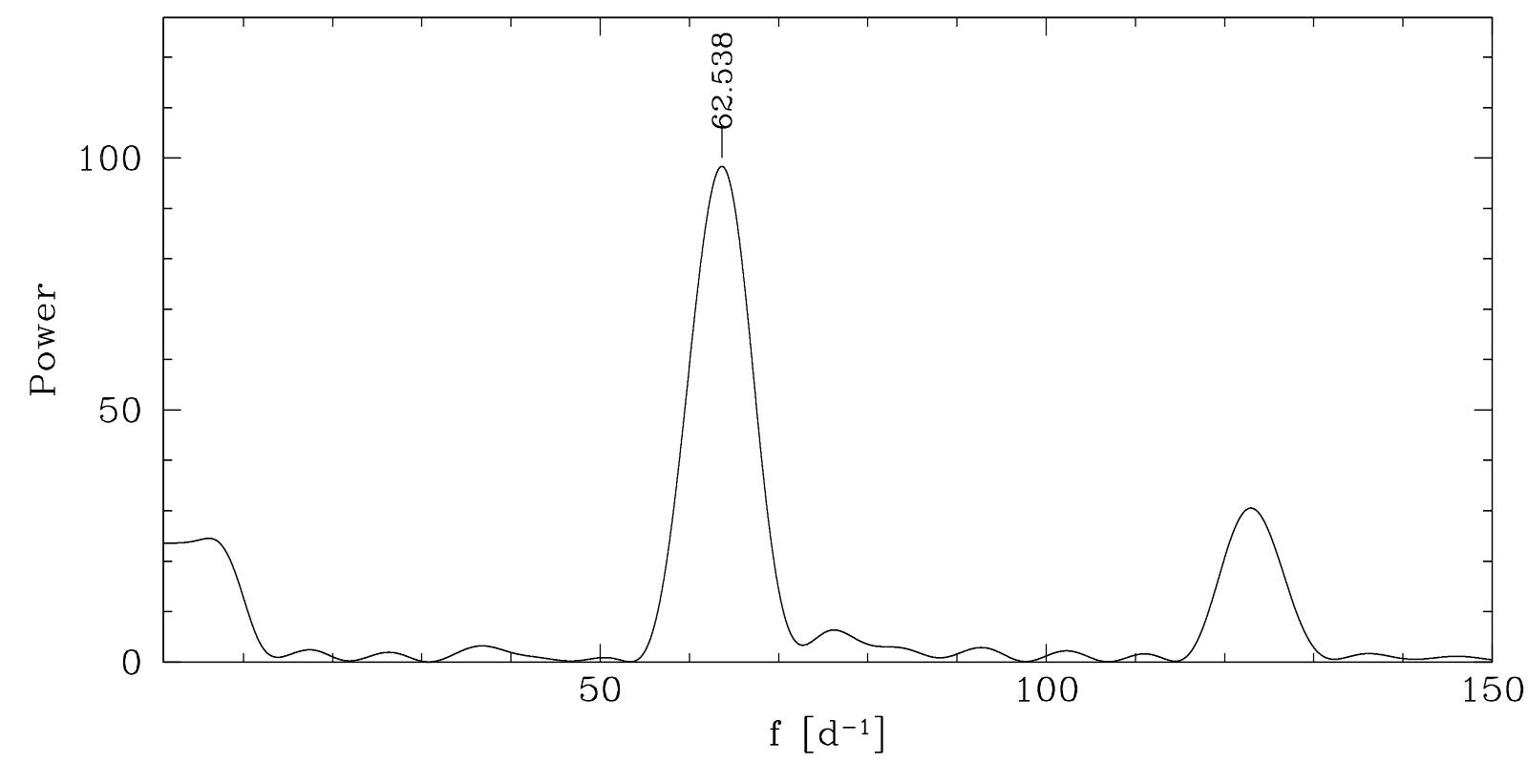}
\includegraphics[width=\columnwidth]{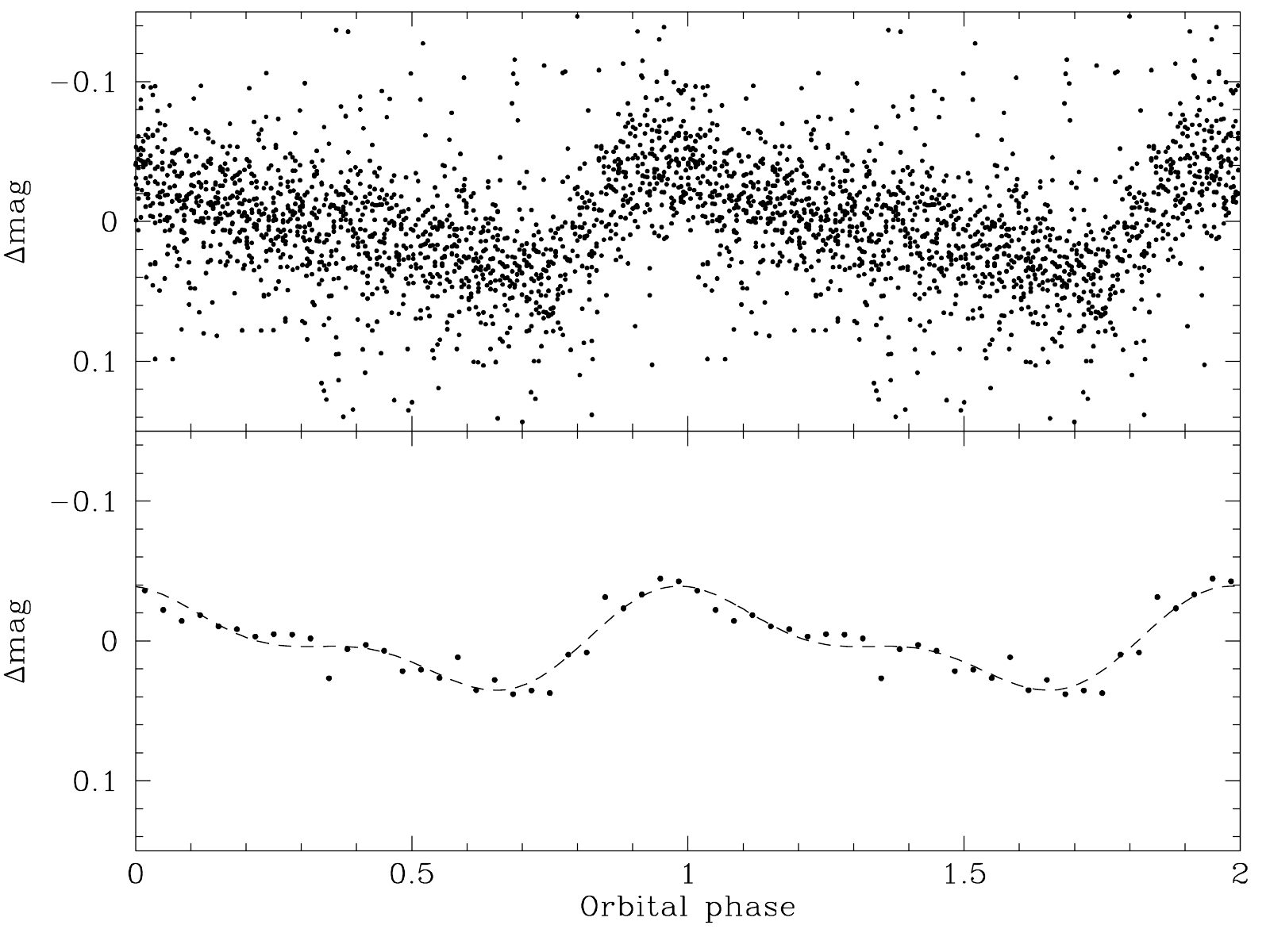}
\end{center}
\caption{\label{f-sdss1831} \textit{Top panel:} The TNT light curve of SDSS\,J1831+4202 obtained on 2023 May 3 shows clear short-period variability. 
\textit{Middle panel:} The Scargle periodogram of these data shows a strong peak at $f=62.538$\,\id. 
\textit{Bottom panels:} The light curves (top: individual data; bottom: data binned into 30 phases) folded with $P=23.026$\,min.}
\end{figure}

We obtained time-series photometry of SDSS\,J1831+4202 during the period April 2023 to May 2024 for a total of 15 nights. The system was at a nearly constant brightness, $17.4-17.6$\,mag, except on 2023 April 28 when we found SDSS\,J1831+4202 in a fainter state at $18.5$\,mag. The system brightened again to $17.1$\,mag by the time of our next observation on 2023 May 3 (Table\,\ref{t-obslog3}). Overall, the light curves obtained in the high state are characterised by short-time scale variation of about $10-30$\,min with an amplitude of $\simeq0.1$\,mag. A clear superhump-like structure was detected in the 2023 May 3 light curve (Fig.\,\ref{f-sdss1831}, top panel). The periodogram of these data contains a strong peak at $f=62.538\pm0.026\,\id$ (Fig.\,\ref{f-sdss1831}, middle panel), or $P=23.026\pm0.097$\,min where the uncertainty was derived from a sine-fit to the data. The phase-folded light curve (Fig.\,\ref{f-sdss1831}, bottom panel) corroborates the superhump-like structure of the photometric variability. Our TNT observations confirm the period of $23.07374\pm0.00001$\,min derived by \citet{kato23-1} from the sparse ZTF data, and securely identify this photometric variability as either the orbital or a superhump signal. SDSS\,J1831+4202 was observed by \textit{TESS} in Sectors 74, 80 and 81, but because of the contamination by a nearby bright star these data were not useful to refine the orbital period of the system.

\section{Discussion and Conclusions} \label{sec-discussion}
The formation and evolution of AM\,CVn stars are not well understood. There are three possible channels describing the formation of these binaries, characterised by the nature of the donor star. In the most compact binaries ($\Porb<10$\,min) the primary white dwarf is thought to accrete materials from another white dwarf \citep{paczynski67-1}, and in longer period systems from a partially degenerate He star \citep{savonijeetal86-1,iben+tutukov87-1}. Some of the longest-period systems in the population may have formed as hydrogen cataclysmic variables, but have lost their hydrogen envelopes through evolution and mass transfer \citep{podsiadlowskietal03-1}. One potential method to discriminate between the various evolutionary models is to assess the chemical composition of the donor star, set mainly by CNO-processing during the evolution \citep{nelemansetal10-1}. The amount of CNO-processing a helium star had undergone by the time it leaves the common envelope to become a mass-transferring AM\,CVn star determines the relative amounts of N and C left in the star, $\mathrm{N/C}>100$ for white dwarf donor models, but $\mathrm{N/C}<10$ for He star donors with $\Porb>20$\,min \citep{nelemansetal10-1}. The most reliable way to measure the donor star abundances is by analysing the composition of the atmosphere of the primary, which is enriched with the accreted material (e.g. \citealt{tolozaetal19-1})~--~however, this requires in practically all cases high-quality ultraviolet spectroscopy. A less robust proxy for the composition of the donor is to investigate the relative strengths of the emission lines from the accretion flow (e.g. \citealt{gaensickeetal03-1}). 

\begin{figure}
\centerline{
\includegraphics[width=\columnwidth]{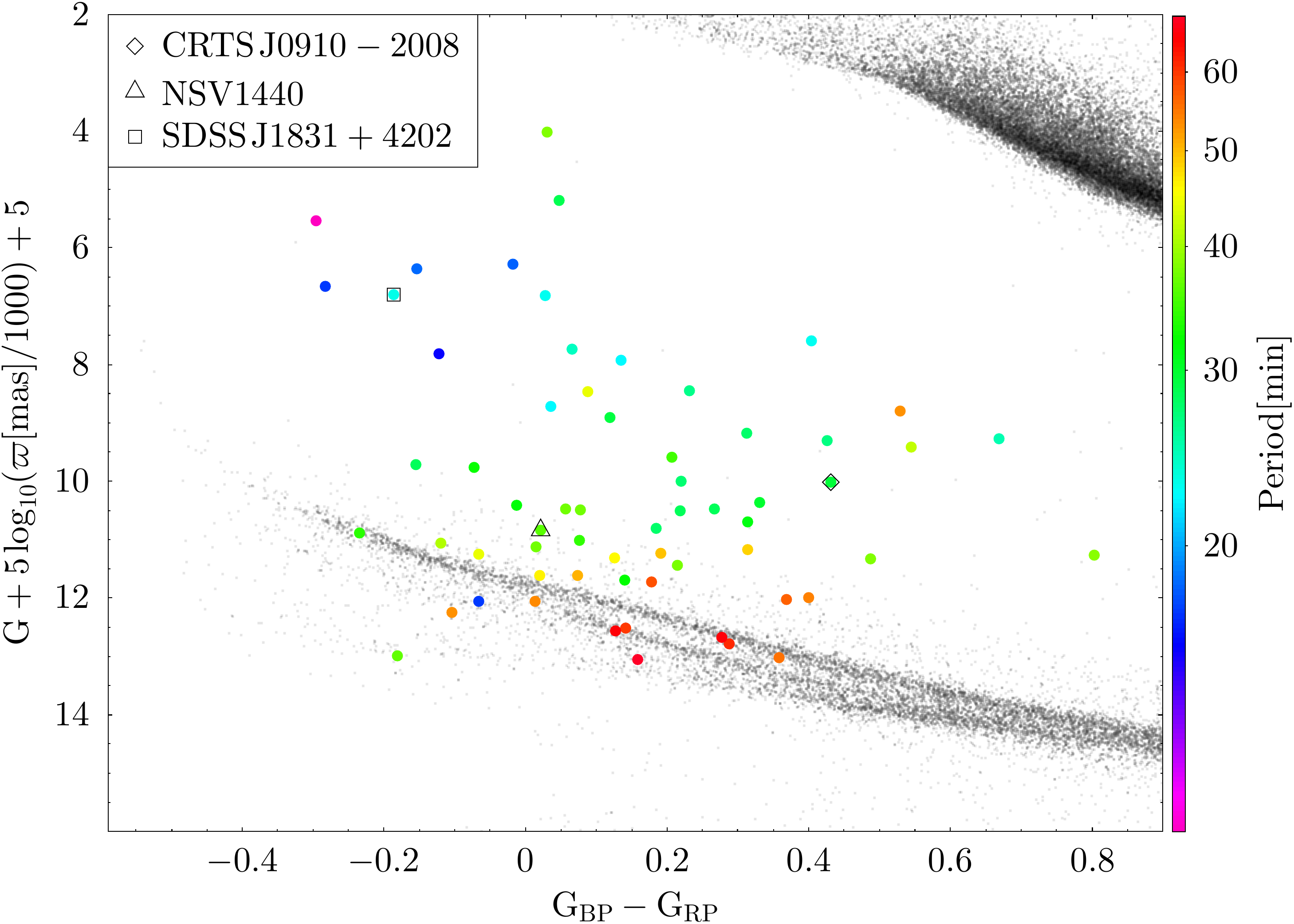}}
\caption{\label{f-hrd} 
\textit{Gaia} Hertzsprung-Russell diagram (HRD) displaying a selection of stars within 100\,pc (gray) and the 63 confirmed AM\,CVn systems (coloured filled circles) with known orbital periods and \textit{Gaia} parallaxes (from \citealt{green23-1}). The newly confirmed systems are highlighted by black outlines SDSS\,J1831+4202 (square), NSV\,1440 (triangle), and CRTS\,J0910$-$2008 (diamond). Systems with short orbital periods and high mass transfer rates tend to be intrinsically luminous and blue, in contrast, the systems with the longest orbital periods and lowest mass transfer rates are firmly located near or within the white dwarf cooling sequence. Systems undergoing dwarf nova outbursts show some scatter in the HRD, as the photometry within \textit{Gaia} DR3 may have sampled different states (see the appendix of \citealt{inightetal22-1}.)}
\end{figure}

The spectrum of CRTS\,J0910$-$2008 is rich in neutral He, N, and Si, as well as ionised Ca and Fe. There is a slight red-ward asymmetry in the He{\sc i}\,5875 line, whose wavelength matches the \Ion{Na}{I} D doublet (e.g. \citealt{yamaguchietal23-1}).  Whereas our spectral range does not include any strong C lines, but the strength of the N lines, as well as the tentative detection of Na suggest that CRTS\,J0910$-$2008 is more likely to have a white dwarf donor than a helium star donor. No N is detected in the spectrum of NSV\,1440, which could indicate a lower rate of conversion of C to N, and it is overall less metal-rich than that of CRTS\,J0910$-$2008, exhibiting only He, Si and weak Ca lines.

Another method to constrain the donor parameters (and hence the evolutionary scenario) is through the superhump period excess \citep{greenetal18-1, greenetal20-1}, if both the orbital and superhump periods are known. 

Adopting the period measured from the long \textit{TESS} observations as being the orbital period, $\Porb=36.56\pm0.03$\,min, and the ``stage A'' superhump period of \citet{isogaietal19-1}, $\Psh=36.98$\,min, we find a period excess of $\epsilon = (\Psh-\Porb)/\Porb = 0.011$, and a mass ratio of $q = M_2/M_1 = 0.032$, using equation\,(5) of \citet{kato+osaki13-1}. Whereas this value is lower than that found by \citet{isogaietal19-1}, it is still consistent with a semi-degenerate donor. 

The superhump period of CRTS\,J0910$-$2008 detected from the two outbursts in 2015 and 2018 is $\Psh=29.7$\,min, but its orbital period is not yet known. Despite being fairly faint in quiescence, $G\simeq19.4$, a spectroscopic determination of the orbital period is feasible thanks to the strong emission lines of this system. We confirm a photometric period of $P=23$\,min for SDSS\,J1831+4202, which may be either the orbital or (more likely) the superhump period. A spectroscopic measurement of \Porb\ will be challenging given the broad and shallow absorption lines in this system. 

We show the location within the Hertzsprung-Russell diagram (HRD) of the 63 AM\,CVn stars that currently have both measured orbital or superhump periods as well as \textit{Gaia} parallaxes \citep{greenetal24-1}.  As expected, SDSS\,J1831+4202 which spends most of the time in a high state with a hot and steady-state disc has the  bluest $G_\mathrm{BP}-G_\mathrm{RP}$ colour. AM\,CVn stars with the longest orbital periods ($\gtrsim50$\,min) tend to cluster near, or within the white dwarf cooling sequence, reflecting the weak contribution of the accretion disc to the overall emission. Systems with frequent outbursts are distributed over a relatively large area of HRD parameter space, in fact, CRTS\,J0910$-$2008 is the reddest system in our study, even though it has a shorter period than NSV\,1440. This somewhat scattered appearance in the HRD was discussed by \citet{inightetal22-1} as a consequence of \textit{Gaia} sampling the systems at random epochs, resulting in average $G$, $G_\mathrm{BP}$ and $G_\mathrm{RP}$ magnitudes that do not necessarily reflect those of their most topical state. Hence, some care needs to be taken when predicting the properties of an AM\,CVn system solely based on its location in the HRD.

\section*{Acknowledgements}
We the thank the referee, Dr. Oliveira Kepler, for a prompt and constructive report.
This research has received funding support from the National Science, Research and Innovation Fund (NSRF) via the Program Management Unit for Human Resources \& Institutional Development, Research and Innovation (Grant No. B05F640046).
This project has received funding from the European Research Council (ERC) under the European Union’s Horizon 2020 research and innovation programme (Grant agreement No. 101020057).
This work has made use of data obtained at the Thai National Observatory on Doi Inthanon, operated by NARIT.
This work has made use of data obtained by the PROMPT-8 telescope, owned by National Astronomical Research Institute of Thailand, and operated by the Skynet Robotic Telescope Network.
The CSS survey is funded by the National Aeronautics and Space Administration under Grant No. NNG05GF22G issued through the Science Mission Directorate Near-Earth Objects Observations Program.  The CRTS survey is supported by the U.S.~National Science Foundation under grants AST-0909182.
Based on observations obtained with the Samuel Oschin 48-inch Telescope at the Palomar Observatory as part of the Zwicky Transient Facility project. ZTF is supported by the National Science Foundation under Grant No. AST-1440341 and a collaboration including Caltech, IPAC, the Weizmann Institute for Science, the Oskar Klein Center at Stockholm University, the University of Maryland, the University of Washington, Deutsches Elektronen-Synchrotron and Humboldt University, Los Alamos National Laboratories, the TANGO Consortium of Taiwan, the University of Wisconsin at Milwaukee, and Lawrence Berkeley National Laboratories. Operations are conducted by COO, IPAC, and UW. Based on observations obtained at the Southern Astrophysical Research (SOAR) telescope, which is a joint project of the Minist\'{e}rio da Ci\^{e}ncia, Tecnologia e Inova\c{c}\~{o}es (MCTI/LNA) do Brasil, the US National Science Foundation’s NOIRLab, the University of North Carolina at Chapel Hill (UNC), and Michigan State University (MSU).
%

\section*{Data Availability}
The ASAS-SN, ATLAS and ZTF light curves can be retrieved from the respective archives, and we will make the TNT light curves available upon reasonable requests communicated to the lead author. 



\bibliographystyle{mnras}

\input{bibliography.tex}



\newpage
\appendix

\section{Observation logs}

\begin{table*}
\caption{Comparison stars used for the differential photometry\label{t-compstars}.}
\setlength{\tabcolsep}{0.95ex}
\begin{flushleft}
\begin{tabular}{lccccccc}
\hline\noalign{\smallskip}
Comparison star & Identifier &  Magnitudes \\\hline
C1 & USNO-A2.0 0675-10057431 & $B=15.5$, $R=15.0$ \\
C2 & USNO-A2.0 0675-10058614 & $B=17.8$, $R=17.0$ \\
C3 & USNO-A2.0 0675-10058349 & $B=13.5$, $R=13.7$ \\
C4 & USNO-A2.0 0675-10062884 & $B=13.9$, $R=14.0$ \\
C5 & USNO-A2.0 0675-10058295 & $B=16.6$, $R=16.0$ \\
C6 & USNO-A2.0 0075-00639610 & $B=12.7$, $R=12.1$ \\
C7 & USNO-A2.0 0075-00642821 & $B=15.3$, $R=16.5$ \\
C8 & USNO-A2.0 0075-00643994 & $B=13.9$, $R=16.0$ \\
C9 & USNO-A2.0 1275-10094237 & $B=16.2$, $R=15.2$ \\\hline
\end{tabular}
\end{flushleft}
\end{table*}

\begin{table*}
\caption{Log of the photometric observations of CRTS\,J0910$-$2008\label{t-obslog}.}
\setlength{\tabcolsep}{0.95ex}
\begin{flushleft}
\begin{tabular}{lccccccc}
\hline\noalign{\smallskip}
Date & UT &  Telescope & Filter & Exp. & Frames  & \multicolumn{1}{c}{Comparison} & Mag.\\
     &    &            &        & (s)  &         & \multicolumn{1}{c}{Star}       &      \\
\hline\noalign{\smallskip}
2015 Feb 12 & 03:46-07:59 & PROMPT8 & V & 30 & 208 & C1 & 16.4 \\            
2015 Feb 13 & 01:13-08:55 & PROMPT8 & V & 45 & 286 & C1 & 15.9 \\
2015 Feb 14 & 01:04-08:56 & PROMPT8 & V & 35 & 355 & C1 & 16.6 \\
2015 Feb 15 & 01:05-07:16 & PROMPT8 & V & 55 & 149 & C1 & 17.5 \\ 
2015 Feb 16 & 02:01-07:59 & PROMPT8 & V & 40 & 217 & C1 & 16.8 \\ 
2015 Feb 17 & 00:33-07:59 & PROMPT8 & V & 100 & 135 & C1 & 18.1 \\ 
2015 Feb 19 & 00:22-09:08 & PROMPT8 & V & 100 & 69 & C1 & 18.6 \\
2015 Feb 22 & 00:19-08:43 & PROMPT8 & V & 100 & 64 & C1  & 18.8 \\
2015 Feb 23 & 00:18-08:54 & PROMPT8 & V & 100 & 61 & C1 & 18.9 \\
2015 Feb 24 & 00:17-08:38 & PROMPT8 & V & 100 & 58 & C1 & 19.0 \\
2015 Feb 25 & 00:16-08:19 & PROMPT8 & V & 100 & 33 & C1 & 19.1 \\
2015 Feb 26 & 00:14-08:27 & PROMPT8 & V & 100 & 37 & C1 & 19.1 \\
2015 Feb 27 & 00:13-00:22 & PROMPT8 & V & 100 & 5 & C1 & $>$19.1 \\
2018 Jan 14 & 16:24-17:39 & TNT & KG5 & 30 & 135 & C2 & 20.3 \\
2018 Feb 22 & 16:50-18:56 & TNT & KG5 & 30 & 221 & C2 & 20.4 \\
2018 Feb 24 & 14:47-14:49 & TNT & KG5 & 30 & 4   & C2 & 19.9  \\
2018 Mar 03 & 14:05-14:05 & TNT & KG5 & 30 & 2   & C2 & 19.1  \\
2018 Mar 04 & 12:33-12:36 & TNT & KG5 & 5, 30 & 13  & C2 & 19.0  \\
2018 Mar 05 & 14:55-14:56 & TNT & KG5 & 30 & 3   & C2 & 19.6  \\
2018 Mar 18 & 12:27-18:01 & TNT & KG5 & 1 & 6431 & C3 & 14.0 \\
2018 Mar 19 & 12:24-18:16 & TNT  & KG5 & 3 & 4530 & C3 & 14.2 \\
2018 Mar 20 & 12:20-18:18 & TNT & KG5 & 3 & 5157 & C3 & 14.3 \\
2018 Mar 22 & 15:47-17:59 & 0.5\,m-TNO & V & 20 & 352 & C3 & 14.8 \\
2018 Mar 23 & 14:07-17:45 & 0.5\,m-TNO & V & 20 & 563 & C3 & 14.9 \\
2018 Mar 24 & 14:42-17:46 & 0.5\,m-TNO & V & 20 & 188 & C3 & 15.8 \\
2018 Mar 27 & 12:38-13:19 & 0.5\,m-TNO & V & 120 & 19 & C3 & 16.8 \\
2018 Mar 28 & 01:02-03:20 & ROAD 0.4\,m & CV & 60 & 82 & C4 & 15.9 \\
2018 Mar 30 & 02:16-04:22 & ROAD 0.4\,m & CV & 60 & 48  &  C4 & 17.3 \\
2018 Mar 31 & 03:57-04:19 & ROAD 0.4\,m & CV & 60 & 10  & C4 & 16.7 \\
2018 Apr 01 & 03:53-04:15 & ROAD 0.4\,m & CV & 60 & 10  & C4 & 16.4 \\
2018 Apr 03 & 00:31-04:07 & ROAD 0.4\,m & CV & 60 & 98  & C4 & 17.5 \\
2018 Apr 04 & 03:40-04:03 & ROAD 0.4\,m & CV & 60 & 9  & C4 & 16.3 \\
2018 Apr 05 & 03:37-03:57 & ROAD 0.4\,m & CV & 60 & 8  & C4 & 18.0 \\
2018 Apr 06 & 03:33-03:53 & ROAD 0.4\,m & CV & 60 & 8  & C4 & 17.9 \\
2018 Apr 06 & 12:45-14:50 & TNT         & KG5 & 15 & 432 & C5 & 17.9 \\
2018 Apr 07 & 03:29-03:49 & ROAD 0.4\,m    & CV & 60 & 8  & C4 & 18.4 \\
2018 Apr 08 & 03:22-03:44 & ROAD 0.4\,m    & CV & 60 & 9  & C4 & 18.4 \\
2018 Apr 08 & 12:32-15:55 & TNT       & KG5 & 15 & 773 & C5 & 18.0 \\
2018 Apr 09 & 03:21-03:41 & ROAD 0.4\,m    & CV & 60 & 8  & C4 & 18.5 \\
2018 Apr 10 & 03:22-03:37 & ROAD 0.4\,m    & CV & 60 & 4  & C4 & 18.6 \\
2021 Apr 01 & 14:38-14:39 & TNT & KG5 & 15 & 5 & C5 & 19.4 \\
2021 Apr 02 & 13:46-13:54 & TNT & KG5 & 15, 30 & 13 & C5 & 19.8 \\
2023 Jan 22 & 19:03-19:26 & TNT & KG5 & 15 & 76 & C5 & 19.2 \\
2023 Mar 10 & 12:36-12:42 & TNT & KG5 & 30 & 13 & C5 & 19.0 \\
2023 Mar 11 & 16:38-16:41 & TNT & KG5 & 15 & 10 & C5 & 19.1 \\
2023 Mar 17 & 13:06-13:08 & TNT & KG5 & 15 & 10 & C5 & 18.9 \\
2023 Apr 10 & 14:35-14:38 & TNT & KG5 & 15 & 10 & C2 & 19.4 \\
2023 Apr 17 & 14:22-14:25 & TNT & KG5 & 15 & 10 & C2 & 19.4 \\
2023 Apr 23 & 14:07-14:09 & TNT & KG5 & 15 & 5 & C2 & 19.1 \\
2023 Apr 24 & 14:04-14:07 & TNT & KG5 & 15 & 10 & C2 & 19.3 \\
2023 Apr 28 & 14:13-14:17 & TNT & KG5 & 15 & 6 & C2 & 18.7 \\
2024 Feb 22 & 17:17-17:20 & TNT & KG5 & 10 & 15 & C5 & 19.0 \\
\noalign{\smallskip}\hline
\end{tabular}
\end{flushleft}
\end{table*}

\begin{table*}
\caption{Log of the photometric observations of NSV\,1440\label{t-obslog2}.}
\setlength{\tabcolsep}{0.95ex}
\begin{flushleft}
\begin{tabular}{lccccccc}
\hline\noalign{\smallskip}
Date & UT &  Telescope & Filter & Exp. & Frames  & \multicolumn{1}{c}{Comparison} & Mag.\\
     &    &            &        & (s)  &         & \multicolumn{1}{c}{Star}       &      \\
\hline\noalign{\smallskip}
2015 Nov 22 & 04:06-07:49 & PROMPT8 & $V$ & 20 & 350 & C6 & 12.4 \\            
2015 Nov 23 & 04:34-07:45 & PROMPT8 & $V$ & 10 & 452 & C6 & 12.5 \\
2015 Nov 24 & 05:00-07:41 & PROMPT8 & $V$ & 10 & 381 & C6 & 12.6 \\
2015 Nov 26 & 05:56-06:49 & PROMPT8 & Clear & 100 & 20 & C7 & 18.7 \\
2015 Nov 27 & 01:41-05:53 & PROMPT8 & Clear & 100 & 100 & C7 & 17.9 \\[1ex]
\underline{Long time series:} &&&&&&&\\
                   & Nr. of nights &&&&&&  Mag range\\
2015 Nov 21 - 2016 Jan 9 & 50 & ROAD 0.4\,m & CV & 30--60 & 3894 & C8 & 12.96--18.13\\
%
2019 Feb 17 - 2019 Apr 25 & 68 & ROAD 0.4\,m & CV & 30 & 2363 & C8 & 13.44--18.04\\[1ex]
\underline{\textit{TESS:}} &&&&&&&\\
                        & Sector & \\
2020 Jul 05 - 2020 Jul 28 & 27 & \textit{TESS} & $T$ & 120 & 13490 & \\
2020 Jul 31 - 2020 Aug 24 & 28 & \textit{TESS} & $T$ & 120 & 14909 & \\
2021 May 27 - 2021 Jun 24 & 39 & \textit{TESS} & $T$ & 120 & 19377 & \\
2023 May 05 - 2023 Jun 01 & 65 & \textit{TESS} & $T$ & 120 & 18968 & \\
2023 Jun 02 - 2023 Jun 25 & 66 & \textit{TESS} & $T$ & 120 & 11777 & \\
2023 Jul 01 - 2023 Jul 25 & 67 & \textit{TESS} & $T$ & 120 & 13661 & \\
2023 Jul 29 - 2023 Aug 22 & 68 & \textit{TESS} & $T$ & 120 & 15334 & \\
%
\hline
\end{tabular}
\end{flushleft}
\end{table*}

\begin{table*}
\caption{Log of the photometric observations of SDSS\,J183131.63+420220.2\label{t-obslog3}.}
\setlength{\tabcolsep}{0.95ex}
\begin{flushleft}
\begin{tabular}{lccccccc}
\hline\noalign{\smallskip}
Date & UT &  Telescope & Filter & Exp. & Frames  & Comparison  & Mag.\\
     &    &            &        & (s)  &         & Star        &      \\
\hline\noalign{\smallskip}
2023 Apr 10 & 18:44-22:34 & TNT & KG5 & 10 & 1223 & C9 & 17.4 \\ 
2023 Apr 24 & 21:02-22:17 & TNT & KG5 & 5 & 752 & C9 & 17.8 \\
2023 Apr 28 & 17:12-19:57 & TNT & KG5 & 10 & 874 & C9 & 18.5 \\
2023 May 03 & 19:34-22:15 & TNT & KG5 & 5 & 1439 & C9 & 17.1 \\
2023 May 04 & 20:47-21:42 & TNT & KG5 & 5 & 366 & C9 & 17.2 \\
2024 Mar 30 & 21:54-22:39 & TNT & KG5 & 10 & 246 & C9 & 17.4 \\
2024 Apr 01 & 21:49-22:41 & TNT & KG5 & 5 & 528 & C9 & 17.4 \\
2024 Apr 17 & 19:54-22:20 & TNT & KG5 & 8 & 969 & C9 & 17.4 \\
2024 Apr 23 & 16:41-22:11 & TNT & KG5 & 5 & 1457 & C9 & 17.4 \\
2024 Apr 25 & 19:11-22:14 & TNT & KG5 & 5 & 1232 & C9 & 17.4 \\
2024 Apr 27 & 19:23-19:47 & TNT & KG5 & 5 & 65 & C9 & 17.5 \\
2024 Apr 29 & 19:23-22:12 & TNT & KG5 & 10 & 774 & C9 & 17.5 \\
2024 May 06 & 18:00-18:24 & TNT & KG5 & 7 & 183 & C9 & 17.6  \\
2024 May 07 & 18:30-18:56 & TNT & KG5 & 7 & 175 & C9 & 17.4 \\
2024 May 12 & 17:52-19:28 & TNT & KG5 & 5 & 939 & C9 & 17.4 \\
%
\hline
\end{tabular}
\end{flushleft}
\end{table*}


\begin{table*}
\caption{Photometric periods detected in the three AM\,CVn stars.\label{t-Psh}} 
\setlength{\tabcolsep}{0.95ex}
\begin{flushleft}
\begin{tabular}{llccc}
\hline\noalign{\smallskip}
Object & Date & $P$\,(min) \\
\hline\noalign{\smallskip}
CRTS\,J0910$-$2008        & 2015 Feb 12 - 16                    & $29.677\pm0.004$ &\\
                          & 2018 Mar 18 - 19                    & $30.073\pm0.009$ &\\
                          & 2018 Mar 20 Mar                     & $30.199\pm0.016$&\\
                          & 2018 Mar 22 - 23                    & $29.722\pm0.007$ &\\
                          & 2015 Feb 12 - 16 + 2018 Mar 22 - 23 & $29.700\pm0.004$\\ 
NSV1440                   & 2015 Nov 22 - 24                    & $37.53\pm0.13$\\
                          & 2019 Feb 17 - 20                    & $36.525\pm0.017$ \\
                          & 2023 May 05 - Jun 01                & $36.6019\pm0.0019$ &\\
                          & 2023 Jun 02 - Jun 25                & $36.5361\pm0.0037$ &\\
                          & 2023 Jul 01 - Jul 25                & $36.5419\pm0.0020$ &\\
                          & 2023 Jul 29 - Aug 22                & $36.5599\pm0.0027$ &\\
SDSS\,J183131.63+420220.2 & 2023 May 03                         & $23.026\pm0.097$ &\\
\hline
\end{tabular}
\end{flushleft}
\end{table*}

\bsp	
\label{lastpage}
\end{document}